\documentclass[12pt]{article}

\usepackage{multirow}
\usepackage{graphicx}
\usepackage{subfig}
\usepackage{xcolor}
\usepackage{ulem}
\usepackage{adjustbox}
\usepackage{pdflscape}

\begin{document}

\title{ Analytic Solutions for Geodesic Motion in Static Axially Symmetric Spacetime}

\author{R. Chan$^1$\thanks{e-mail: chan@on.br}, M.F.A. da Silva$^2$\thanks{e-mail: mfasnic@gmail.com}, and N.O. Santos$^3$\thanks{e-mail: nilton.santos@obspm.fr}\\
{\small $^1$ Coordena\c{c}\~ao de Astronomia e Astrof\'{\i}sica, Observat\'orio Nacional,}\\
{\small rua General Jos\'e Cristino 77, S\~ao Crist\'ov\~ao, 20921-400, Rio de Janeiro, RJ, Brazil.}\\
{\small $^2$ Departamento de F\'{i}sica Te\'{o}rica, Universidade do Estado do Rio de Janeiro (UERJ),}\\
{\small  Rio de Janeiro, RJ 20550-900, Brazil.}\\
{\small $^3$ Sorbonne Universit\'e, UPMC Universit\'e Paris 06, LERMA, Observatoire de Paris-Meudon,}\\
{\small 5 place Jules Janssen, F-92195 Meudon Cedex France.}}
\maketitle

\begin{abstract}
A procedure to find static axially symmetric solutions to the Einstein field equations is presented. We obtained two general solutions and five particular solutions,
which depend on the existence conditions for circular and $z$ direction motion.
Our endeavour consists  making a thoroughrowly analysis of all the possible geodesics solutions stemming from this spacetime.
\end{abstract}

\section{Introduction}

 Perturbations on the spherically symmetric vacuum spacetime arising from a perturbation of its source distribution may produce highly unexpected results stemming from its deep non linear character \cite{Herrera2007}. Examples are like directional singularities \cite{Cooperstock1974}, depending on the way that one approaches the source to detect its existence;
 and the appearance of repulsive gravitational forces \cite{Herrera2007}. Furthermore, these results can lead to sources where its gravitational field diminishes until it disappears producing, at the end, no effect on the motion of particles \cite{Herrera1999}. Another interesting property suggesting some links to these non Newtonian results where shown produces by stationary cylindrically symmetric fields  evolving into repulsive ones  \cite{Herrera1998}.
 
 In this same vein, the stationary black hole, given by the Kerr metric, was shown to produce repulsive fields in certain regions of its ergosphere  \cite{de Felice1992}.
 
 These non Newtonian results have in common axially symmetric deformations of static or stationary fields or both of them. The physical reason for the existence of these non Newtonian repulsive fields is still unknown. Nevertheless, even with this lack of knowledge it is suggested by a number of researchers, that extragalactic jets might be produced or partly produced by these repulsive fields \cite{Gariel2014} by the galaxy 
  M87.
  
 The observed data for the jet produced by the galaxy  M87, which one of the most studied, fit to a high degree of approximation the shell structure predicted by the Kerr spacetime and the observed energy radiated by its highly collimated jet.
 
 In this  scenario, so far, it seems important to study these peculiar kind of phenomena undergoing relativistic  gravitational  fields from its mathematical point of view as well as astrophysical implication.
 
 With this context in mind we propose here to undertake the general study of geodesics in a static axially symmetric spacetime.
 
In Section 2 we present the field equations and in Section 3 we show one particular solution of these field equations. In Section 4 and 5 we show two general solutions of the field equations. In Section 6 we present the general geodesics expressions in a static axially symmetric spacetime. In Section 7 we show the circular geodesics equations. In Section 8 we present the geodesics in z-direction equations. In Section 9 we show the geodesics in z-$\phi$ direction equations. In Section 10 we present the geodesics in $\rho$ direction equations. In Section 11 we present the conditions necessary to find particular solutions of the field equations. In the Sections 12, 13 and 14 we show five new solutions (3, 4, 5(a), 5(b) and 5(c)), using the conditions obtained in Section 11. The rest of Sections 15, 16, 17 and 18 we present the study of the geodesics of all found solutions.
In Section 19 we show our Conclusions.
 
\section{The Field Equations}

We consider an axially symmetric  static spacetime with a metric element
given in Weyl coordinates \cite{Islam1985}\cite{Scott1986}\cite{Hernandez-Pastora2016}
\begin{equation}
ds^2=-fdt^2+e^{g}(d{\rho}^2+dz^2)+ld\phi^2, \label{ds2}
\end{equation}
where $f$, $g$ and $l$ are all functions of ${\rho}$ and $z$. To represent  cylindrical symmetry, we impose the following ranges on the coordinates
\begin{equation}
-\infty\leq t\leq\infty, \;\; {\rho} \geq 0, \;\; -\infty<z<\infty, \;\; 0\leq\phi\leq 2\pi.
\end{equation}
We number the coordinates $x^0=t$, $x^1={\rho}$, $x^2=z$ and $x^3=\phi$. 

The field equations  in the vacuum are given by
 
\begin{eqnarray}
G_{00}&=&-\frac{1}{4{ l ^{2}{e}^{g }}} \left\{f  \left( 2  g_{,zz} l ^{2}+2  g_{,\rho\rho} l ^{2}+2  l_{,zz}l +2  l_{,\rho\rho} l - l_{,\rho} ^{2}- l_{,z} ^{2}\right) \right\}=0,\label{G00}\\
G_{11}&=&\frac{1}{4{l ^{2} f ^{2}}} \left\{ g_{,\rho}  f_{,\rho}  l ^{2}f - g_{,z}  f_{,z}  l ^{2}f + g_{,\rho}  l_{,\rho} l  f ^{2}- g_{,z}  l_{,z} l  f ^{2}+2  l_{,zz} l f ^{2}+\right.\nonumber\\
&&\left. 2  f_{,zz}  l ^{2}f - f_{,z} ^{2} l ^{2}+l  l_{,\rho} f f_{,\rho} +l  l_{,z} f f_{,z} - l_{,z} ^{2} f ^{2}\right\}=0,\label{G11}\\
G_{12}&=&-\frac{1}{4{ l ^{2} f ^{2}}} \left\{- g_{,\rho}  f_{,z}  l ^{2}f - g_{,z}  f_{,\rho}  l ^{2}f - g_{,z}  l_{,\rho} l  f ^{2}- g_{,\rho}  l_{,z} l  f ^{2}+\right.\nonumber\\
&&\left. 2  l_{,z\rho} l  f ^{2}+2  f_{,z\rho}  l ^{2}f - f_{,\rho}  f_{,z}  l ^{2}- l_{,\rho} l_{,z}  f ^{2}\right\}=0,\label{G12}\\
G_{22}&=&-\frac{1}{4{ l ^{2} f ^{2}}} \left\{ g_{,\rho}  f_{,\rho}  l ^{2}f - g_{,z}  f_{,z}  l ^{2}f + g_{,\rho}  l_{,\rho} l  f ^{2}- g_{,z}  l_{,z} l  f ^{2}-\right.\nonumber\\
&&\left. 2  f_{,\rho\rho}  l ^{2}f + f_{,\rho} ^{2} l ^{2}-l  l_{,\rho} f f_{,\rho} -2  l_{,\rho\rho}l  f ^{2}-l  l_{,z} f f_{,z} + l_{,\rho} ^{2} f ^{2}\right\},\label{G22}\\
G_{33}&=&\frac{1}{4{ f ^{2}{e}^{g }}} \left\{l  \left( 2  g_{,zz}  f ^{2}+2  g_{,\rho\rho}  f ^{2}+2  f_{,zz} f +2  f_{,\rho\rho} f - f_{,\rho} ^{2}- f_{,z} ^{2} \right) \right\}=0,\label{G33}
\end{eqnarray}

where the commas in the subscript stand for differentiation of the indeces.

In order to integrate the field equations, we will assume first that
\begin{eqnarray}
D^2=f l,
\end{eqnarray}
thus we have
 
\begin{eqnarray}
2De^{g}  G_{00} &=& -2f \left( D_{,{\rho}{\rho}} + D_{,zz} \right) + D \left(f_{,{\rho}{\rho}} + f_{,zz} 
\right) - f D \left( g_{,{\rho}{\rho}} + g_{,zz} \right)-\nonumber\\
&&\left( D_{,{\rho}} f_{,{\rho}} + D_{,z} f_{,z} \right) +
\frac{3f}{2D} \left( f_{,{\rho}} l_{,{\rho}} + f_{,z} l_{,z} \right)=0,\label{G00a}\\
2D G_{11} &=& 2D_{,zz} + D_{,{\rho}} g_{,{\rho}} - D_{,z} g_{,z} +
\frac{1}{2D}\left(  f_{,{\rho}} l_{,{\rho}} - f_{,z} l_{,z} \right)=0,\label{G11a}\\
2D G_{12} &=& -2D_{,{\rho}z} + D_{,{\rho}} g_{,z} + D_{,z} g_{,{\rho}} +
\frac{1}{2D}\left( f_{,{\rho}} l_{,z} + f_{,z} l_{,{\rho}} \right)=0,\label{G12a}\\
2D  G_{22} &=& 2D_{,{\rho}{\rho}} - D_{,{\rho}} g_{,{\rho}} + D_{,z} g_{,z} -
\frac{1}{2D}\left(  f_{,{\rho}} l_{,{\rho}}  - f_{,z} l_{,z} \right)=0,\label{G22a}\\
2D e^{g}  G_{33} &=& 2l\left( D_{,{\rho}{\rho}} + D_{,zz} \right) - D \left(l_{,{\rho}{\rho}} + l_{,zz} \right) + 
lD \left( g_{,{\rho}{\rho}} + g_{,zz} \right)+\nonumber\\
&&D_{,{\rho}} l_{,{\rho}} + D_{,z} l_{,z} -
\frac{3l}{2D} \left( f_{,{\rho}} l_{,{\rho}}  + f_{,z} l_{,z} \right)=0.\label{G33a}
\end{eqnarray}

From the field equations (\ref{G11a}) and (\ref{G22a}) we can write
\begin{eqnarray}
D_{,{\rho}{\rho}}+D_{,zz}=0,\label{Drr}
\end{eqnarray}
which is a Laplace equation of the quantity $D$. It is possible to
demonstrate without loss of generality that $D_{,{\rho}{\rho}}=D_{,zz}=0$.
Integrating $D_{,{\rho}{\rho}}=0$ twice we obtain that
\begin{eqnarray}
D^2=fl={\rho}^2,\label{flr2}
\end{eqnarray}
rescaling the two arbitrary integration constants that appear, without loss
of generality.

\section{$\gamma$ Metric Solution}

This solution , given by the $\gamma$ metric, is a particular axially symmetric solution studied with the Erez-Rosen spherical coordinates \cite{Erez1959}\cite{Herrera2000}.
 This is given in terms of the metric by 
\cite{Esposito1975} 
\begin{equation}
ds^2=-e^{2\lambda}dt^2+e^{2\mu-2\lambda}(d{\rho}^2+dz^2)+e^{-2\lambda}{\rho}^2 d\phi^2, \label{ds2t}
\end{equation}
\begin{eqnarray}
\lambda &=& \frac{\gamma}{2} \ln\left(\frac{r_1+r_2-2 m}{r_1+r_2+2 m}\right), \\
\mu &=& \frac{\gamma^2}{2} \ln\left[\frac{(r_1+r_2+2 m)(r_1+r_2-2 m)}{4 r_1 r_2}\right], \\
\end{eqnarray}
where
\begin{eqnarray} 
r_1 &=& \sqrt{({\rho}^2+(z-m)^2}, \\ 
r_2 &=& \sqrt{({\rho}^2+(z+m)^2},
\end{eqnarray}
and where the density  of mass $\gamma$ is distributed symmetrically along the axis 
for a length $2m$ \cite{Esposito1975}.
Making the suitable coordinate transformation, we get thus
\begin{eqnarray}
f &=& e^{2 \lambda},\\
l &=& {\rho}^2/f, \\
g &=& 2 (\mu-\lambda),
\end{eqnarray}
 with
\begin{eqnarray}
&&f=\left( {\frac {r_a+r_b-2 m}{r_a+r_b+2 m}} \right) ^{\gamma},\nonumber\\
&&g= \ln \left[ \left( \frac{1}{2} {\frac {-{m}^{2}+{\rho }^{2}+r_a r_b+{z}^{2}}{{r_a} r_b}} \right) ^{{\gamma}^{2}} \left( {\frac {r_a
		+r_b-2 m}{r_a+r_b+2 m}} \right) ^{-\gamma} \right],\nonumber\\
&&l= {\rho}^{2} /f,\label{gamma}
\end{eqnarray}
where ${r_a}=\sqrt {{m}^{2}-2 z m+{\rho}^{2}+{z}^{2}}$ and 
${r_b}=\sqrt {{m}^{2}+2 z m+{\rho}^{2}+{z}^{2}}$.

Substituting these last equations into (\ref{G00})-(\ref{G33})  
we can see that they are all fulfilled.

In the next two sections we will substitute the equation (\ref{flr2}) into the field equations (\ref{G00})-(\ref{G33})  (since the equations (\ref{G00a})-(\ref{G33a}) are used only as auxiliary ones in order to integrate once)
and resolve the partial differential equation system using the Maple 16 software.
Thus, we can get two general solutions that will presented below. \\

\section{General Solution 1}

\begin{eqnarray}
f&=&c_2 e^{c_1 z},\nonumber\\
l&=&\rho^2/f,\nonumber\\
g&=&-c_1 z-\frac{1}{4} c_1^2 {\rho}^2+c_3.\label{sol1}
\end{eqnarray}\\

 Note that the only way to obtain a solution independent of the $z$
coordinate is by setting the constant $c_1=0$. This would lead us directly to 
the Minkowski spacetime, that is, this solution does not admit the Levi-Civita spacetime
as a particular case.

\section{General Solution 2}

\begin{eqnarray}
f &=& d_1 d_4  e^{\frac{1}{4} d_0 {\rho}^2-\frac{1}{2} d_0 z^2-d_3 z} {\rho}^{-d_2},\nonumber\\
l&=&\rho^2/f,\nonumber\\
g &=& \left(\frac{1}{2} d_2^2+ d_2\right) \ln({\rho})+d_0^2 \left[\frac{1}{32} ({\rho}^4-8 {\rho}^2 z^2)\right] +\nonumber\\
&&\frac{d_0}{4} \left[(-2 d_3 z- d_2-1) {\rho}^2+2 z^2 (1+d_2)\right] -\nonumber\\
&&\frac{1}{4}d_3^2 {\rho}^2 +d_3 (1+d_2) z+d_5,\label{sol2}
\end{eqnarray}
where $c_1$, $c_2$, $c_3$, $d_0$, $d_1$, $d_2$, $d_3$, $d_4$ and $d_5$ are arbitrary
constants of integration. If $d_0=d_3=0$ and $d_2=-4\sigma$  we get the 
Levi-Civita spacetime in General Relativity, where $\sigma$ is the linear energy density.

\section{General Geodesics Equations}

The general geodesics equations using the metric (\ref{ds2})
are given by
\begin{eqnarray}
\frac{d^2x^{\alpha}}{d \tau^2}+\Gamma_{\beta \delta}^{\alpha} \frac{dx^\beta}{d \tau} \frac{dx^\delta}{d \tau}=0,
\end{eqnarray}
which can be written as
\begin{eqnarray}
&&\ddot \rho- {\frac { g_{,\rho}  \dot z ^{2}{e}^{g }- g_{,\rho}  \dot \rho ^{2}{e}^{g }-2  g_{,z}  \dot \rho  \dot z {e}^{g }- f_{,\rho}  \dot t ^{2}+ l_{,\rho}  \dot \phi ^{2}}{2{e}^{g }}}=0,\label{ddr}\\
&&\ddot z + {\frac {2  g_{,\rho}  \dot \rho  \dot z {e}^{g }+ g_{,z}  \dot z ^{2}{e}^{g }- g_{,z}  \dot \rho ^{2}{e}^{g }+ f_{,z}  \dot t ^{2}- l_{,z}  \dot \phi ^{2}}{2{e}^{g }}}=0,\label{ddz}\\
&&\ddot \phi +{\frac { \dot \phi  \left(  l_{,z} \dot z +l_{,\rho} \dot \rho  \right) }{l }}=0,\label{ddphi}\\\
&&\ddot t +{\frac { \dot t \left(  f_{,z} \dot z + f_{,\rho} \dot \rho  \right) }{f }}=0.\label{ddt}
\end{eqnarray}

Comparing these geodesic equations with those presented in \cite{Islam1985} we can notice
that they agree completely.

\section{Circular Geodesic Equations}
Let us assume that
\begin{eqnarray}
{\rho}={\rho}_0, \dot {\rho}=0, \ddot {\rho}=0, z=z_0, \dot z=0, \ddot z=0,
\end{eqnarray}
where $\rho_0$ and $z_0$ are constants.
Then from equations (\ref{ddr})-(\ref{ddt}) we have that
\begin{eqnarray}
&&  {\frac {  f_{,\rho}  \dot t ^{2}- l_{,\rho}  \dot \phi ^{2}}{2{e}^{g }}}=0,\label{ddr1}\\
&&  {\frac {  f_{,z}  \dot t ^{2}- l_{,z}  \dot \phi ^{2}}{2{e}^{g }}}=0,\label{ddz1}\\
&&\ddot \phi=0,\label{ddphi1}\\\
&&\ddot t =0.\label{ddt1}
\end{eqnarray}

From the equations (\ref{ddr1})-(\ref{ddt1}) we can already notice that, if
$f_{,{\rho}}$, $f_{,z}$, $l_{,{\rho}}$ and $l_{,z}$ are different from zero in general
them the unique solution is $\phi(\tau)=constant$ and $t(\tau)=constant$ , implying the inexistence of circular geodesics.

\section{Geodesic Equations in z-Direction}

Let us assume that
\begin{eqnarray}
{\rho}={\rho}_0, \dot {\rho}=0, \ddot {\rho}=0, \dot \phi=0, \ddot \phi=0,
\end{eqnarray}
then from equations (\ref{ddr})-(\ref{ddt}) we have that
\begin{eqnarray}
&&{\frac { -g_{,\rho}  \dot z ^{2}{e}^{g }+ f_{,\rho}  \dot t ^{2}}{2{e}^{g }}}=0,\label{ddr2}\\
&&\ddot z + {\frac { g_{,z}  \dot z ^{2}{e}^{g }+ f_{,z}  \dot t ^{2}}{2{e}^{g }}}=0,\label{ddz2}\\
&&\ddot t +{\frac {   f_{,z} \dot t \dot z   }{f }}=0.\label{ddt2}
\end{eqnarray}
where the equation (\ref{ddphi}) is identically zero.

 From equation (\ref{ddz2}) we can see that the acceleration depends on how the metric functions $f$ and $g$ depend on the $z$ coordinate. However, in a special case, if $ g_{,z}  \dot z ^{2}{e}^{g }=- f_{,z}  \dot t ^{2}$ we can show that we do not have acceleration in the $z$ direction.

\section{Geodesic Equations in z-{\bf $\phi$}-Direction}

Let us assume that
\begin{eqnarray}
{\rho}={\rho}_0, \dot {\rho}=0, \ddot {\rho}=0,
\end{eqnarray}
then from equations (\ref{ddr})-(\ref{ddt}) we have that
\begin{eqnarray}
&&- {\frac { g_{,\rho}  \dot z ^{2}{e}^{g }- f_{,\rho}  \dot t ^{2}+ l_{,\rho}  \dot \phi ^{2}}{2{e}^{g }}}=0,\label{ddr3}\\
&&\ddot z + {\frac { g_{,z}  \dot z ^{2}{e}^{g }+ f_{,z}  \dot t ^{2}- l_{,z}  \dot \phi ^{2}}{2{e}^{g }}}=0,\label{ddz3}\\
&&\ddot \phi +{\frac { l_{,z} \dot \phi    \dot z    }{l }}=0,\label{ddphi3}\\
&&\ddot t +{\frac {  f_{,z} \dot t  \dot z    }{f }}=0.\label{ddt3}
\end{eqnarray}

\section{Geodesic Equations in {\bf $\rho$}-Direction}

Let us assume that
\begin{eqnarray}
{z}={z}_0, \dot {z}=0, \ddot {z}=0, \dot \phi=0, \ddot \phi=0,
\end{eqnarray}
then from equations (\ref{ddr})-(\ref{ddt}) we have that
\begin{eqnarray}
&&\ddot \rho + {\frac {  g_{,\rho}  \dot \rho ^{2}{e}^{g }+ f_{,\rho}  \dot t ^{2}}{2{e}^{g }}}=0,\label{ddr4}\\
&&  {\frac {- g_{,z}  \dot \rho ^{2}{e}^{g }+ f_{,z}  \dot t ^{2}}{2{e}^{g }}}=0,\label{ddz4}\\
&&\ddot t +{\frac { f_{,\rho} \dot t   \dot \rho  }{f }}=0.\label{ddt4}
\end{eqnarray}
where the equation (\ref{ddphi}) is identically zero.

\section{Conditions for Particular Solutions}

We can observe from circular geodesics equations (\ref{ddr1})-(\ref{ddz1})
that we only have circular orbit if and only if (\ref{ddr1}) or (\ref{ddz1}) are identically zero. Thus, we can get particular solutions if we assume the following conditions below to solve the field equations (\ref{G00a})-(\ref{G33a}), thus
\begin{eqnarray}
f_{,z}=0,\label{condfz}\\
l_{,z}=0,\label{condlz}
\end{eqnarray}
or
\begin{eqnarray}
f_{,\rho}=0,\label{condfrho}\\
l_{,\rho}=0.\label{condlrho}
\end{eqnarray}

We can also observe from $z$-direction geodesics equations (\ref{ddr2})-(\ref{ddz2})
that can get particular solutions if we assume the following conditions
to solve the field equations (\ref{G00a})-(\ref{G33a}), thus
\begin{eqnarray}
g_{,\rho}=0,\label{condgrho}\\
g_{,z}=0,\label{condgz}\\
f_{,\rho}=0,\\
f_{,z}=0.
\end{eqnarray}
 Note that last two conditions are identical to (\ref{condfrho})-(\ref{condlrho}).  Thus, these conditions will be
considered using (\ref{condfz})-(\ref{condgz}).

In the next Section we will solve the field equations (\ref{G00a})-(\ref{G33a})
using the  all the conditions (\ref{condfz})-(\ref{condgz})
independent of the type of geodesic studied.

\section{Solution 3 with Circular Geodesic}

This solution is based in the two conditions (\ref{condfz})-(\ref{condlz}), thus
\begin{eqnarray}
g&=&{\frac{1}{2} C_1  \left( C_1-2 \right) \ln(\rho) } + C_3,\nonumber\\
f&=&{\rho}^{C_1}C_2,\nonumber\\
l&=&\rho^2/f.\label{sol3}
\end{eqnarray}

\section{Solution 4 with Circular Geodesics}

This solution is based in the two conditions (\ref{condfrho})-(\ref{condlrho}), thus
\begin{eqnarray}
g&=&-C_1 z-\frac{1}{4} C_1^2 \rho^2+C_3,\nonumber\\
f&=&{e^{C_1 z}}C_2,\nonumber\\
l&=&\rho^2/f.\label{sol4}
\end{eqnarray}

\section{Solutions 5(a), 5(b) and 5(c) with Geodesic in z-Direction}

These solutions are based in the conditions (\ref{condgrho}) and (\ref{condgz}), 
thus we get three different solutions. Note that the conditions (\ref{condfz}) and (\ref{condfrho}) are used already for obtaining Solutions 3 and 4.

\subsection{Solution 5(a) with Geodesic in z-Direction}

\begin{eqnarray}
g&=&C_1,\nonumber\\
f&=&C_2 {\rho}^{2},\nonumber\\
l&=&\rho^2/f.\label{sol5a}
\end{eqnarray}
 
This solution resembles the Rindler metric but the only difference is the metric
function $g_{\phi\phi}=1/C_2$ which it should be $\rho^2$.

\subsection{Solution 5(b) with Geodesics in z-Direction}

\begin{eqnarray}
g&=&C_2 \ln(\rho) + C_1,\nonumber\\
f&=&C_3 {\rho}^{1+\sqrt {1+2 C_2}},\nonumber\\
l&=&\rho^2/f.\label{sol5b}
\end{eqnarray}
 
Note that this solution is the Levi-Civita  solution assuming that $C_1=a$,
$C_2=4\sigma(2\sigma-1)$ and $C_3=a$, where $a$ and $\sigma$ are constants.

\subsection{Solution 5(c) with Geodesics in z-Direction}

\begin{eqnarray}
g&=&\frac{1}{2} C_1 \rho^2-\frac{1}{2} \ln(\rho)+ C_2,\nonumber\\
f&=&{e^{\sqrt {C_1} \left( C_3-z \right) \sqrt {2}}}\rho,\nonumber\\
l&=&\rho^2/f.\label{sol5c}
\end{eqnarray}

In the next Sections we will present the analysis of the
circular geodesic, geodesic in the $z$ direction, geodesic in the $z$ and $\phi$ 
direction and the geodesic in $\rho$ direction for the $\gamma$ metric, General Solution 1, General Solution 2, Solution 3, Solution 4, Solutions 5(a), 5(b) and 5(c). 
We will  study all the above solutions since the conditions 
(\ref{condfrho})-(\ref{condgz}) are very similar in all these geodesics.

\section{Analysis of the Circular Geodesic}
\subsection{{\bf $\gamma$} Metric Solution}
Substituting the  equations (\ref{gamma}) into (\ref{ddr1})-(\ref{ddt1}) we get
we obtain that
\begin{eqnarray}
&&\frac {2r}{r_a r_b \left[ 4 {m}^{2}- \left( r_a+r_b \right) ^{2} \right] } \left( - {\frac {{m}^{2}-\rho_0^{2}-r_a r_b-{z}^{2}}{2r_ar_b}} \right) ^{-{\gamma}^{2}} \times\nonumber\\
&&\left\{ -m\dot t ^{2}\gamma  \left( r_a+r_b \right)  \left( {\frac {r_a+r_b-2 m}{r_a+r_b+2 m}} \right) ^{2 \gamma}+\right.\nonumber\\
&&\left. \left[ {m}^{4}+ \left( 2 \rho_0^{2}-r_a r_b-2 {z}^{2}\right) {m}^{2}-\rho_0^{2}\gamma  \left( r_a+r_b \right) m+ \right.\right.\nonumber\\
&&\left.\left.\left( \rho_0^{2}+{z}^{2} \right)  \left( \rho_0^{2}+r_a r_b+{z}^{2} \right)  \right]  \dot \phi ^{2} \right\} =0,\label{geocirc1}\\
&&- \frac {2\gamma m \left[ \left( r_a-r_b \right) m+z\left( r_a+r_b \right)  \right] }{r_ar_b \left( 4 {m}^{2}-\left( r_a+r_b \right) ^{2} \right) } \left( - {\frac {{m}^{2}-\rho_0^{2}-r_a r_b-{z}^{2}}{2r_ar_b}} \right) ^{-{\gamma}^{2}} \times\nonumber\\
&&\left[ \rho_0^{2} \dot \phi ^{2}+ \dot t ^{2} \left( {\frac {r_a+r_b-2 m}{r_a+r_b+2 m}} \right) ^{2 \gamma} \right] =0,\label{geocirc2}\\
&&{\ddot \phi}=0,\label{geocirc3}\\
&&{\ddot t}=0.\label{geocirc4}
\end{eqnarray}
Solving the equations (\ref{geocirc1})-(\ref{geocirc4}) simultaneously 
for any value of $z$ we have that
\begin{eqnarray}
\phi(\tau)&=&constant,\\
t(\tau)&=&constant,
\end{eqnarray}
meaning that we do not have circular geodesic. However,  if we study the circular orbit in the $z=0$ plane, we can see that the equation (\ref{geocirc2}) is identically zero (since $r_a=r_b$ in $z=0$).
Thus, solving the equations (\ref{geocirc1}) and (\ref{geocirc3})-(\ref{geocirc4})
simultaneously we get
\begin{eqnarray}
&&t \left( \tau \right) =C_2 \tau+C_3,\\
&&\phi \left( \tau \right) =\pm \frac {t \left( \tau \right) }{ \left( -m\gamma+\sqrt {{m}^{2}+\rho_0^{2}} \right) r}\times\nonumber\\
&&\sqrt { \left( -m\gamma+\sqrt {{m}^{2}+\rho_0^{2}} \right)  \left( {\frac {\sqrt {{m}^{2}+\rho_0^{2}}-m}{\sqrt {{m}^{2}+\rho_0^{2}}+m}} \right) ^{2 \gamma}m\gamma}+C_1
,
\end{eqnarray}
where $C_1$, $C_2$, $C_3$, $C_4$, $C_5$, $C_6$, $C_7$ and $C_8$, hereinafter, are constants of integration.
We calculate the frequency of the circular orbit as
\begin{eqnarray}
w^2={\frac {m\gamma}{\rho_0^{2} \left( -m\gamma+\sqrt {{m}^{2}+\rho_0^{2}}\right) } \left( {\frac {\sqrt {{m}^{2}+\rho_0^{2}}-m}{\sqrt {{m}^{2}+\rho_0^{2}}+m}} \right) ^{2 \gamma}}.
\end{eqnarray}

Notice that this result is not  directly comparable to that obtained in a previous paper \cite{Herrera1999} since the authors used a spherical Erez-Rosen \cite{Erez1959} coordinate transformation in order to analyze the circular geodesic.
\subsection{General Solution 1}
Substituting the  equations (\ref{sol1}) into (\ref{ddr1})-(\ref{ddt1}) we get
\begin{eqnarray}
&&-{\frac {{e^{\frac{1}{4} {c_1}^{2}{\rho_0}^{2}-c_3}} 
{\dot \phi} ^{2}\rho_0 }{{c_2}}}=0,\\
&&\frac{1}{2} {\frac {c_1  \left( {e^{2 c_1 z+\frac{1}{4} {c_1}^{2}{\rho_0}^{2}-c_3}}{c_2}^{2} 
{\dot t} ^{2}+{e^{\frac{1}{4} {c_1}^{2}{\rho_0}^{2}-c_3}} 
{\dot \phi} ^{2}{\rho_0 }^{2} \right) }{c_2}}=0,\\
&&{\ddot \phi}=0,\\
&&{\ddot t}=0. 
\end{eqnarray}
Solving these equations simultaneously for any value of $z$, even for $z=0$, we have that
\begin{eqnarray}
\phi(\tau)&=&constant,\\
t(\tau)&=&constant,
\end{eqnarray}
meaning that we do not have circular geodesic.
\subsection{General Solution 2}
Substituting the  equations (\ref{sol2}) into (\ref{ddr1})-(\ref{ddt1}) we get
\begin{eqnarray}
&&\frac{1}{4{d_1 d_4}} \left\{ -2 {d_1}^{2}{d_4}^{2} \dot t ^{2} \left( \rho_0^{-\frac{1}{2} {d_2}^{2}-1-2 d_2}d_2-\frac{1}{2} \rho_0^{-\frac{1}{2} {d_2}^{2}+1-2 d_2}d_0 \right) \right.\nonumber\\
&&\left. {e^{-\frac{1}{32} {d_0}^{2}\rho_0^{4}+\frac{1}{32}  \left( 8 {d_0}^{2}{z}^{2}+16 d_0 d_3 z+8 d_0 d_2+8 {d_3}^{2}+16 d_0 \right) \rho_0^{2}-\frac{1}{2} d_0  \left( d_2+2 \right) {z}^{2}-d_3  \left( d_2+2 \right) z-d_5}}-\right.\nonumber\\
&&\left. 2  \left(  \left( d_2+2 \right) \rho_0^{-\frac{1}{2} {d_2}^{2}+1}-\frac{1}{2} \rho_0^{-\frac{1}{2} {d_2}^{2}+3}d_0\right)  \dot \phi ^{2}\right.\nonumber\\
&&\left. {e^{-\frac{1}{32} {d_0}^{2}\rho_0^{4}+\frac{1}{32}  \left( 8 {d_0}^{2}{z}^{2}+16 d_0 d_3 z+8 d_0 d_2+8 {d_3}^{2} \right) \rho_0^{2}-\frac{1}{2} d_0 d_2 {z}^{2}-d_3 zd_2-d_5}} \right\}=0,\\
&&-\frac{1}{2{d_1 d_4}} \left\{ \left( d_0 z+d_3 \right)  \left[  \dot t ^{2}\rho_0^{-\frac{1}{2} d_2  \left( d_2+4 \right) }{d_1}^{2}{d_4}^{2}e^{-\frac{1}{32} {d_0}^{2}\rho_0^{4}} \right.\right.\times\nonumber\\
&&\left.\left. e^{\frac{1}{32}  \left( 8 {d_0}^{2}{z}^{2}+16 d_0 d_3 z+8 d_0 d_2+8 {d_3}^{2}+16 d_0 \right) \rho_0^{2}-\frac{1}{2} d_0  \left( d_2+2\right) {z}^{2}-d_3  \left( d_2+2 \right) z-d_5}+\right.\right.\nonumber\\
&&\left.\left. \rho_0^{-\frac{1}{2} {d_2}^{2}+2} \dot \phi ^{2}{e^{-\frac{1}{32} {d_0}^{2}\rho_0^{4}+\frac{1}{32}  \left( 8 {d_0}^{2}{z}^{2}+16 d_0 d_3 z+8 d_0 d_2+8 {d_3}^{2} \right) \rho_0^{2}-\frac{1}{2} d_0 d_2 {z}^{2}-d_3 zd_2-d_5}} \right] \right\}=0,\nonumber\\
\\
&&{\ddot \phi}=0,\\
&&{\ddot t}=0\label{eqcircsol2}. 
\end{eqnarray}
Solving  the equations (\ref{eqcircsol2}) simultaneously for any value of $z$,
 even for $z=0$, 
we have  two possible solutions, thus

\begin{eqnarray}
&&\phi_1(\tau)=\pm  \left\{ \sqrt{ {e^{-\frac{1}{32} {d_0}^{2}\rho_0^{4}+\frac{1}{4} {d_0}^{2}\rho_0^{2}{z}^{2}+\frac{1}{2} d_0 d_3 \rho_0^{2}z+\frac{1}{4} d_0 d_2 \rho_0^{2}-\frac{1}{2} d_0 d_2 {z}^{2}}}}\times\nonumber\right.\\
&&\left. \sqrt{ e^{\frac{1}{4} \rho_0^{2}{d_3}^{2}-d_3 zd_2-d_5}\left( -2 \rho_0^{2 d_2+2}d_2+\rho_0^{2 d_2+4}d_0-4 \rho_0^{2 d_2+2} \right)} \right.\times\nonumber\\
&&\left. \sqrt{ {e^{-\frac{1}{32} {d_0}^{2}\rho_0^{4}+\frac{1}{4} {d_0}^{2}\rho_0^{2}{z}^{2}+\frac{1}{2} d_0 d_3 \rho_0^{2}z+\frac{1}{4} d_0 d_2 \rho_0^{2}-\frac{1}{2} d_0 d_2 {z}^{2}+\frac{1}{4} \rho_0^{2}{d_3}^{2}-d_3 zd_2}}} \right.\times\nonumber\\
&&\left.\sqrt{ e^{-d_5+\frac{1}{2} d_0 \rho_0^{2}-d_0 {z}^{2}-2 d_3 z}}\left( d_0 \rho_0^{2}-2 d_2 \right) d_4 d_1 \left( C_2 \tau+C_3 \right) \right\}\times\nonumber \\
&&\left\{ {e^{-\frac{1}{32} {d_0}^{2}\rho_0^{4}+\frac{1}{4} {d_0}^{2}\rho_0^{2}{z}^{2}+\frac{1}{2} d_0 d_3 \rho_0^{2}z+\frac{1}{4} d_0 d_2 \rho_0^{2}-\frac{1}{2} d_0 d_2 {z}^{2}+\frac{1}{4} \rho_0^{2}{d_3}^{2}-d_3 zd_2-d_5}}\right\}^{-1} \times\nonumber\\
&&\left\{\left( -2 \rho_0^{2 d_2+2}d_2+\rho_0^{2 d_2+4}d_0-4 \rho_0^{2 d_2+2}\right) \right\}^{-1}+C_1=0,\\
&&t_1 \left( \tau \right) =C_2 \tau+C_3,
\end{eqnarray}
and
\begin{eqnarray}
\phi_2(\tau)&=&constant,\\
t_2(\tau)&=&constant.
\end{eqnarray}
However, substituting these two real solutions into the equations (\ref{eqcircsol2}) 
we notice that
only the last one give us all the equations (\ref{eqcircsol2}) identically zero.
 This is due to the fact we have a system of quadratic differential equations 
where some of roots might not fulfill completely the original equations. 
Thus, again we do not have circular geodesic.
\subsection{Solution 3}
Substituting the  equations (\ref{sol3}) into (\ref{ddr1})-(\ref{ddt1}) we get
\begin{eqnarray}
&&\frac{1}{2} {\frac {{\rho_0}^{-\frac{1}{2} {C_1}^{2}+2 C_1-1} {\dot t}  ^{2}C_1 {C_2}^{2}+{\rho_0}^{-\frac{1}{2} {C_1}^{2}+1} {\dot \phi}  ^{2} \left( C_1-2 \right) }{e^{C_3} C_2}}=0,\\
&&{\ddot \phi}=0,\\
&&{\ddot t}=0, 
\end{eqnarray}
 with
\begin{eqnarray}
t \left( \tau \right)&=&C_4 \tau+C_5,\\
\phi(\tau)&=&\pm {\frac {\sqrt {- \left( C_1-2\right) {\rho_0}^{2 C_1}C_1}C_2 t \left( \tau \right) }{ \left( C_1-2 \right) r}}+C_3.
\end{eqnarray}

The frequency of circular geodesic is given by
\begin{eqnarray}
\omega^2&=&-{\frac {{\rho_0}^{-2+2 C_1}{C_2}^{2}C_1}{C_1-2}},
\end{eqnarray}
which it is  real if $0<C_1<2$.

\subsection{Solution 4}

Substituting the  equations (\ref{sol4}) into (\ref{ddr1})-(\ref{ddt1}) we get
\begin{eqnarray}
&&-{\frac {{e^{\frac{1}{4} {C_1}^{2}{\rho_0}^{2}}} {\dot \phi}  ^{2}\rho_0}{e^{C_3} C_2}}=0,\\
&&\frac{1}{2} {\frac {C_1  \left( {e^{\frac{1}{4} C_1  \left( C_1 {\rho_0}^{2}+8 z \right) }} {\dot t}  ^{2}{C_2}^{2}+{e^{\frac{1}{4} {C_1}^{2}{\rho_0}^{2}}} {\dot \phi}  ^{2}{\rho_0}^{2} \right) }{e^{C_3} C_2}}=0,\\
&&{\ddot \phi}=0,\\
&&{\ddot t}=0  
\end{eqnarray}

Solving these last equations simultaneously we have that
\begin{eqnarray}
\phi(\tau)&=&constant,\\
t(\tau)&=&constant,
\end{eqnarray}
thus, we do not have  any circular orbit.

\subsection{Solution 5(a)}

Substituting the  equations (\ref{sol5a}) into (\ref{ddr1})-(\ref{ddt1}) we get
\begin{eqnarray}
&&{\frac { {\dot t}  ^{2}C_2 \rho_0}{e^{C_1}}}=0,\\
&&{\ddot \phi}=0,\\
&&{\ddot t}=0.
\end{eqnarray}

Solving these last equations simultaneously we have that
\begin{eqnarray}
\phi(\tau)&=&constant,\\
t(\tau)&=&constant,
\end{eqnarray}
thus, we again do not have  any circular orbit.

\subsection{Solution 5(b)}
Substituting the  equations (\ref{sol5b}) into (\ref{ddr1})-(\ref{ddt1}) we get
\begin{eqnarray}
&&\frac{e^{-C_1}}{2{C_3}} \left[  {\dot \phi}  ^{2} \left( \sqrt {1+2 C_2}-1 \right) {\rho_0}^{-C_2-\sqrt {1+2 C_2}}+\right.\nonumber\\
&&\left. {\rho_0}^{-C_2+\sqrt {1+2 C_2}} {\dot t}  ^{2}{C_3}^{2} \left( 1+\sqrt {1+2 C_2}\right) \right] =0,\\
&&{\ddot \phi}=0,\\
&&{\ddot t}=0
\end{eqnarray}

Solving these last equations simultaneously we have that
\begin{eqnarray}
t \left( \tau \right) &=&C_5 \tau+C_6,\\
\phi(\tau)&=&\pm {\frac{C_3  \left( C_5 \tau+C_6 \right) }{\sqrt {1+2 C_2}-1}} \times\nonumber\\
&&\sqrt {- \left( \sqrt {1+2 C_2}-1 \right) {\rho_0}^{2 \sqrt {1+2 C_2}} \left( 1+\sqrt {1+2 C_2} \right) }+C_4,
\end{eqnarray}

The frequency of circular geodesic is given by
\begin{eqnarray}
\omega^2&=&-2 {\frac {{\rho_0}^{2 \sqrt {1+2 C_2}}C_1 {C_3}^{2}}{ \left( \sqrt {1+2 C_2}-1 \right) ^{2}}},
\end{eqnarray}
which it is  real if $C_1<0$ and $C_2>-\frac{1}{2}$.

\subsection{Solution 5(c)}

Substituting the  equations (\ref{sol5c}) into (\ref{ddr1})-(\ref{ddt1}) we get
\begin{eqnarray}
&&\frac{1}{2}  \sqrt {\rho_0} \left(  {\dot t}  ^{2}{e^{2 \sqrt {-C_1}\sqrt {2}C_3}}-{e^{2 \sqrt {-C_1}\sqrt {2}z}}{\dot \phi}  ^{2} \right) \times\nonumber\\
&&{e^{-\sqrt {2} \left( z+C_3 \right) \sqrt {-C_1}-\frac{1}{2} C_1 {\rho_0}^{2}-C_2}}=0,\\
&&-\frac{1}{2}  \sqrt {2}{\rho_0}^{\frac{3}{2}}\sqrt {-C_1} \left(  {\dot t}  ^{2}{e^{2 \sqrt {-C_1}\sqrt {2}C_3}}+{e^{2 \sqrt {-C_1}\sqrt {2}z}} {\dot \phi}  ^{2} \right)\times\nonumber\\
&& {e^{-\sqrt {2} \left( z+C_3 \right) \sqrt {-C_1}-\frac{1}{2} C_1 {\rho_0}^{2}-C_2}}=0,\\
&&{\ddot \phi}=0,\\
&&{\ddot t}=0  
\end{eqnarray}

Solving these last equations simultaneously we get
\begin{eqnarray}
\phi(\tau)&=&constant,\\
t(\tau)&=&constant,
\end{eqnarray}
Thus, again we do not have  any circular geodesic.

\section{Analysis of the Geodesic in z-Direction}

\subsection{{\bf $\gamma$} Metric Solution}
 In this case we calculated first for $\rho\neq 0$. Since we could not 
find an analytical solution then we have chosen $\rho=0$ for the sake of simpleness for the
presentation of the equations.
Substituting the  equations (\ref{gamma}) into (\ref{ddr2})-(\ref{ddt2}) for $\rho_0=0$ we get
\begin{eqnarray}
&&\frac {2}{\sqrt {r_{a0}}\sqrt {r_{b0}} \left( {m}^{2}-r_{a0} r_{b0}- z ^{2}\right)  \left[ \left(\sqrt {r_{a0}}+\sqrt {r_{b0}}\right)^2-4 m^2 \right]  } \times\nonumber\\
&&\left\{ \frac {m \dot t ^{2} \left[ \left( z +m \right) \sqrt {r_{b0}}-\sqrt {r_{b0}} \left( -z +m\right)  \right]  \left[ {m}^{2}-r_{a0} r_{b0}- z ^{2} \right]  }{\sqrt {r_{a0}}+\sqrt {r_{b0}}+2 m} \right.\times\nonumber\\
&&\left. \left( \sqrt {r_{a0}}+\sqrt {r_{a0}}-2 m \right) {e}^{- {\frac { \left( -{m}^{2}+r_{a0} r_{b0}+ z ^{2} \right)  \left( r_{a0}+r_{b0}+2 m \right) }{2r_{a0} r_{b0}  \left( r_{a0}+r_{b0}-2 m\right) }}}+\right.\nonumber\\
&&\left. \left[ {\frac { \dot z ^{2} \left( r_{a0}+r_{b0}+2 m\right)  \left( -{m}^{2}+r_{a0} r_{b0}+ z^{2} \right) z }{2r_{a0} r_{b0}  \left( r_{a0}+r_{b0}-2 m \right) }}- \right.\right.\nonumber\\
&&\left.\left. \ddot z  \left( {m}^{2}-r_{a0} r_{b0}- z ^{2}\right)  \right] \right. \times\nonumber\\
&&\left. \left[ \sqrt {r_{b0}} \left( {m}^{2}- z ^{2} \right) \sqrt {r_{a0}}-{m}^{4}+2 {m}^{2} z ^{2}- z ^{4} \right]  \right\} =0,\\
&&- \left(  \ddot t  \left[ {m}^{2}- z ^{2} \right] \sqrt {r_{b0}}-2 m \dot t  \dot z  \left( z +m \right)  \right) \sqrt {r_{a0}}+\nonumber\\
&& \left( -{m}^{4}+2 {m}^{2} z ^{2}- z ^{4} \right) \ddot t +2 m \dot t  \dot z  \left( -z +m \right) \sqrt {r_{b0}} \times\nonumber\\
&&\left\{\sqrt {r_{a0}}\sqrt {r_{b0}} \left( \sqrt {r_{a0}}\sqrt {r_{b0}}-{m}^{2}+ z ^{2} \right) \right\}^{-1}=0,
\end{eqnarray}
where $r_{a0}=r_a(\rho_0=0)$ and $r_{b0}=r_b(\rho_0=0)$.

Solving these last equations  simultaneously we can note that we cannot obtain an analytical solution.

\subsection{General Solution 1}
\begin{figure}[!htp]
	\centering	
	\includegraphics[width=6.5cm]{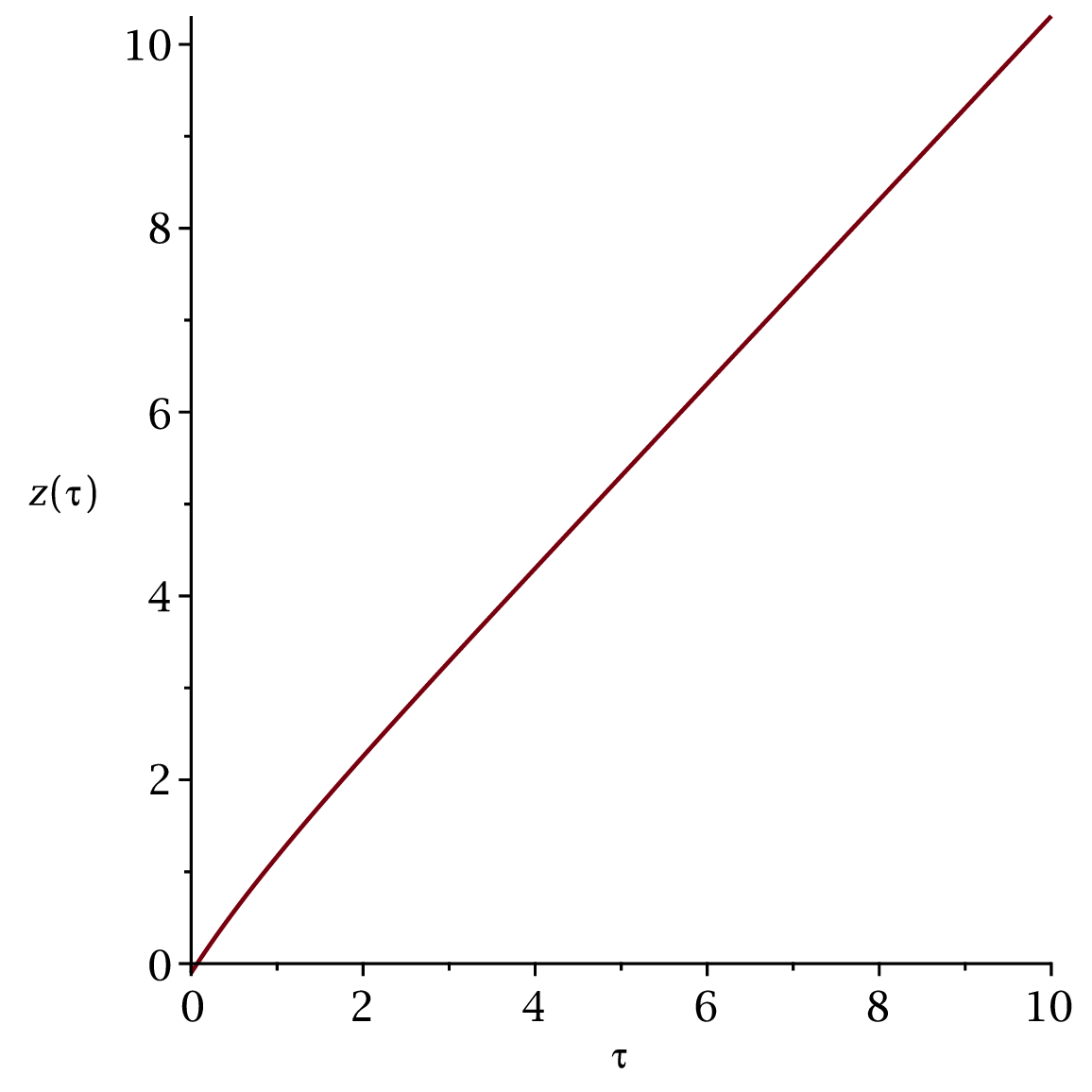}
	\includegraphics[width=6.5cm]{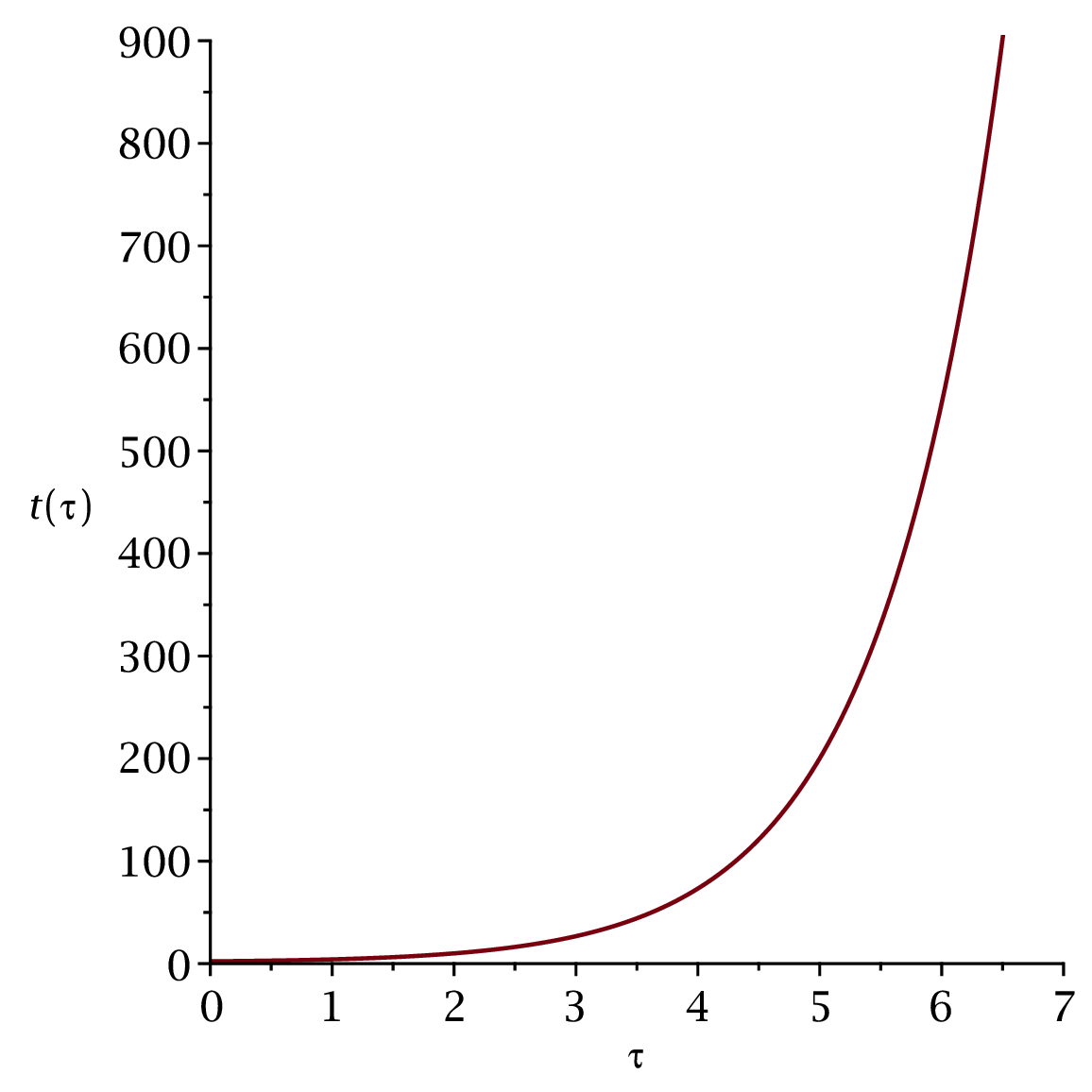}
	\includegraphics[width=6.5cm]{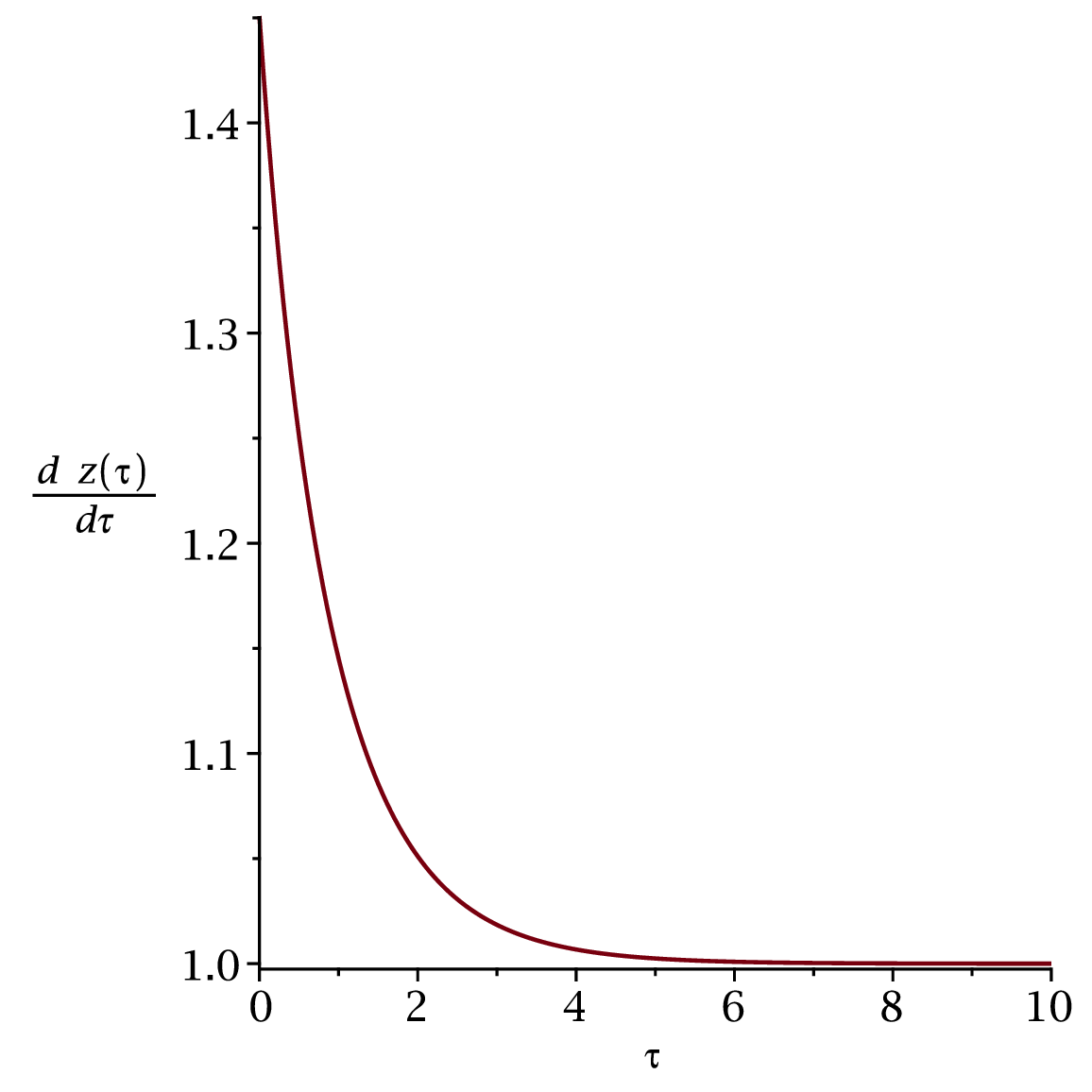}
	\includegraphics[width=6.5cm]{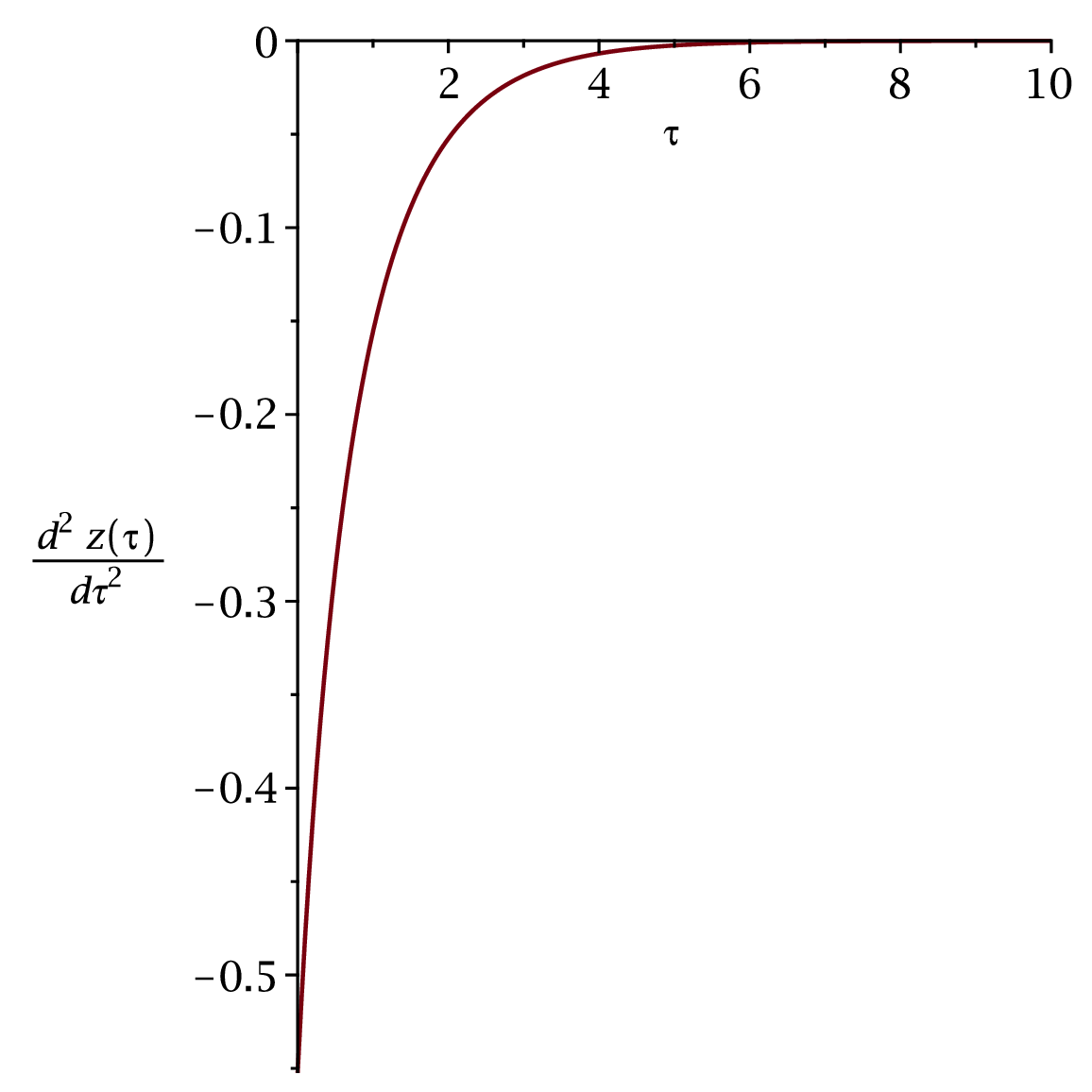}
	\caption{These figures show the time evolution of the $z$-geodesics of the Solution 1
	and its velocity and acceleration of $z(\tau)$ at the axis ${\rho_0}=0$.
	We assume in these figures that $c_1=-1$, $c_2=1$, $c_3=0$, $C_1=1$, $C_2=1$, $C_3=1$
	and $C_4=1$.}
	\label{fig1}
\end{figure}

Substituting the  equations (\ref{sol1}) into (\ref{ddr2})-(\ref{ddt2}) we get
\begin{eqnarray}
&&\frac{1}{4}  {\dot z} ^{2}{c_1}^{2}{\rho}_{{0}}=0,\nonumber\\
&&{\ddot z} +\frac{1}{2} {\frac { {e}^{2c_1 z } {\dot t} ^{2}c_1 c_2}{{e}^{-\frac{1}{4} {c_1}^{2}{{\rho}_{{0}}}^{2}}{e}^{c_3}}}-\frac{1}{2}  {\dot z} ^{2}c_1=0,\nonumber\\
&&{\ddot t} + {\dot t}  {\dot z} c_1=0,\label{geoz}
\end{eqnarray}
For the Solution 1, assuming that ${\rho}_0 \neq 0$, we have that
solving simultaneously the equations (\ref{geoz}) we get

\begin{eqnarray}
z(\tau)&=&constant,\\
t(\tau)&=&constant,
\end{eqnarray}

meaning that we do not have $z$-direction geodesic.
However if we assume that ${\rho}_0=0$ then we have that
solving simultaneously the equations (\ref{geoz}) we obtain four possible solutions
\begin{eqnarray}
t_1(\tau)&=&C_3,\\
z_1(\tau)&=&-2 {\frac {\ln  \left( -\frac{1}{2} C_1 \tau c_1-\frac{1}{2} C_2 c_1 \right) }{c_1}},\\
t_2(\tau)&=&\frac{1}{2} C_1 {\tau}^{2}+C_2 \tau+C_3,\\
z_2(\tau)&=&{\frac {1}{2c_1}\ln  \left[ -{\frac {{C_1}^{2}}{ \left( C_1 \tau+C_2 \right) ^{4}c_2 {c_1}^{2}}} \right] }+\frac{1}{2} {\frac {c_3}{c_1}},\\
t_3(\tau)&=&\frac{1}{6} C_1 {\tau}^{3}+\frac{1}{2} C_2 {\tau}^{2}+C_3 \tau+C_4,\\
z_3(\tau)&=&{\frac {1}{2c_1} \left\{ 4 \ln  \left( 2 \right) +\ln \left[ {\frac {2 C_1 C_3-{C_2}^{2}}{ \left( C_1 {\tau}^{2}+2 C_2 \tau+2 C_3 \right) ^{4}c_2 {c_1}^{2}}} \right] +c_3 \right\} },\\
t_4(\tau)&=&-\frac{1}{8} {{C_1}^{4}{C_2}^{2} {e^{{-\frac {\tau}{C_1}}}} {e^{{-\frac {C_3}{C_1}}}} }-\frac{1}{2} C_2 \tau {C_1}^{2}+\frac{1}{2} {C_1}^{2}{e^{{\frac {\tau}{C_1}}}}{e^{{\frac {C_3}{C_1}}}}+C_4,\\
z_4(\tau)&=&{\frac {1}{8c_1} \left\{ 6 \ln  \left( 2 \right) +\ln \left[ {\frac {1}{{C_1}^{4}{c_1}^{2}c_2}{e^{{\frac {2 \tau+2 C_3}{C_1}}}} \left( C_2 C_1-2 {e^{{\frac {\tau+C_3}{C_1}}}} \right) ^{-4}} \right] +c_3 \right\} },\nonumber\\
\end{eqnarray}
where $C_1$, $C_2$, $C_3$ and $C_4$ are arbitrary constants of integration.
Only the fourth solution $t_4(\tau)$ and $z_4(\tau)$ is physically possible 
 because it is the unique solution that  has fulfilled the equations (\ref{geoz}) identically.
 Again, as explained before in the subsection of the circular geodesics for the General Solution 2, this is due to the fact we have a system of quadratic or biquadratic differential equations 
where some of roots might not fulfill completely the original equations.
We can notice that we do have an accelerated $z$-Direction geodesics at the axis ${\rho}=0$ 
(see Figures \ref{fig1}). 
 
We can notice in the fourth figure of the panel has the same physical characteristics
of relativistic jets highly energetic phenomena as described in \cite{Herrera2007}.
We can see a positive acceleration of a test particle moving along $z$ direction at the beginning.
After sometime, we note that the test particle begins to decelerate. Finally, the
test particle stops the acceleration and it continues to travel at constant velocity.
In that paper, Herrera \& Santos, have interpreted this initial acceleration due
to an existence of a repulsive force. However, in the present work we can clearly see that it is due only to the geometry of the spacetime.

\clearpage

\subsection{Solution 2}

Substituting the  equations (\ref{sol2}) into (\ref{ddr2})-(\ref{ddt2}) we get
 
\begin{eqnarray}
&&-\frac{1}{4} {\frac { \dot z ^{2}{d_2}^{2}}{\rho_0}}-\frac{1}{2} {\frac { \dot z ^{2}d_2}{\rho_0}}-\frac{1}{16}  \dot z ^{2}{d_0}^{2}{\rho_0}^{3}+\frac{1}{4}  \dot z ^{2}{d_0}^{2} \rho_0{z}^{2}+\frac{1}{2}  \dot z ^{2}d_0 d_3 \rho_0z+\nonumber\\
&&\frac{1}{4}  \dot z ^{2}d_0 d_2 \rho_0+\frac{1}{4}  \dot z ^{2}d_0 \rho_0+\frac{1}{4}  \dot z ^{2}\rho_0{d_3}^{2}+\frac{1}{4}  { \dot t ^{2}d_0 \rho_0{e^{\frac{1}{4} d_0 \rho_0^{2}}}{e^{-\frac{1}{2} d_0 {z}^{2}}}d_1 d_4} \times\nonumber\\
&&\left\{e ^{\frac{1}{2} \ln  \left( \rho_0 \right) {d_2}^{2}+\ln  \left( \rho_0 \right) d_2}e ^{\frac{1}{32} {d_0}^{2}{\rho_0}^{4}-\frac{1}{4} {d_0}^{2}{\rho_0}^{2}{z}^{2}-\frac{1}{2} d_0 d_3 {\rho_0}^{2}z-\frac{1}{4} d_0 d_2 {\rho_0}^{2}} \right.\times\nonumber\\
&&\left. e ^{\frac{1}{2} d_0 d_2 {z}^{2}-\frac{1}{4} d_0 {\rho_0}^{2}+\frac{1}{2} d_0 {z}^{2}-\frac{1}{4} {\rho_0}^{2}{d_3}^{2}+d_3 zd_2+d_3 z+d_5}{e^{d_3 z}}{\rho_0}^{d_2} \right\}^{-1}-\nonumber\\
&&\frac{1}{2}  { \dot t ^{2}{e^{\frac{1}{4} d_0 {\rho_0}^{2}}}{e^{-\frac{1}{2} d_0 {z}^{2}}}d_1 d_4 d_2} \times\nonumber\\
&&\left\{ e ^{\frac{1}{2} \ln \left( \rho_0 \right) {d_2}^{2}+\ln \left( \rho_0 \right) d_2+\frac{1}{32} {d_0}^{2}{\rho_0}^{4}-\frac{1}{4} {d_0}^{2}{\rho_0}^{2}{z}^{2}-\frac{1}{2} d_0 d_3 {\rho_0}^{2}z-\frac{1}{4} d_0 d_2 {\rho_0}^{2}} \right.\times\nonumber\\
&&\left. e ^{\frac{1}{2} d_0 d_2 {z}^{2}-\frac{1}{4} d_0 {\rho_0}^{2}+\frac{1}{2} d_0 {z}^{2}-\frac{1}{4} {\rho_0}^{2}{d_3}^{2}+d_3 zd_2+d_3 z+d_5}{e^{d_3 z}}{\rho_0}^{d_2}\rho_0\right\}^{-1}=0,\\
&&\ddot z -\frac{1}{4} \dot z ^{2}{d_0}^{2}{\rho_0}^{2}z -\frac{1}{4}  \dot z ^{2}d_0 d_3 {\rho_0}^{2}+\frac{1}{2}  \dot z ^{2}d_0 d_2 z +\frac{1}{2}  \dot z ^{2}d_0 z \nonumber\\
&&+\frac{1}{2}  \dot z ^{2}d_3 d_2+\frac{1}{2}  \dot z ^{2}d_3-\frac{1}{2} {\dot t ^{2}{e^{\frac{1}{4} d_0 {\rho_0}^{2}}}d_0 z {e^{-\frac{1}{2} d_0  z ^{2}}}d_1 d_4}\times\nonumber\\
&& \left\{ e ^{\frac{1}{2} \ln  \left( \rho_0 \right) {d_2}^{2} +\ln  \left( \rho_0 \right) d_2 + \frac{1}{32} {d_0}^{2}{\rho_0}^{4} -\frac{1}{4} {d_0}^{2}{\rho_0}^{2} z ^{2}-\frac{1}{2} d_0 d_3 {\rho_0}^{2}z -\frac{1}{4} d_0 d_2 {\rho_0}^{2}+\frac{1}{2} d_0 d_2  z ^{2}}\times\right.\nonumber\\
&&\left. e ^{-\frac{1}{4} d_0 {\rho_0}^{2}+\frac{1}{2} d_0  z ^{2}-\frac{1}{4} {\rho_0}^{2}{d_3}^{2}+d_3 z d_2+d_3 z +d_5}{e^{d_3 z }}{\rho_0}^{d_2} \right\}^{-1}-\nonumber\\
&&\frac{1}{2}  {\dot t ^{2}{e^{\frac{1}{4} d_0 {\rho_0}^{2}}}{e^{-\frac{1}{2} d_0  z ^{2}}}d_1 d_4 d_3} \times\nonumber\\
&&\left\{ e ^{\frac{1}{2} \ln \left( \rho_0 \right) {d_2}^{2}+\ln \left( \rho_0 \right) d_2+\frac{1}{32} {d_0}^{2}{\rho_0}^{4}-\frac{1}{4} {d_0}^{2}{\rho_0}^{2} z ^{2}-\frac{1}{2} d_0 d_3 {\rho_0}^{2}z -\frac{1}{4} d_0 d_2 {\rho_0}^{2}+\frac{1}{2} d_0 d_2  z ^{2}} \right.\times\nonumber\\
&&\left.  e ^{-\frac{1}{4} d_0 {\rho_0}^{2}+\frac{1}{2} d_0 z ^{2}-\frac{1}{4} {\rho_0}^{2}{d_3}^{2}+d_3 z d_2+d_3 z +d_5}{e^{d_3 z }}{\rho_{{0}}}^{d_2}\right\}^{-1}=0,\\
&&{\ddot t}  -{\dot t}  {\dot z}  d_0 z - {\dot t}   {\dot z}  d_3=0
\end{eqnarray}

Solving these last equations  simultaneously we  could not obtain an analytical solution.

\subsection{Solution 3}

Substituting the  equations (\ref{sol3}) into (\ref{ddr2})-(\ref{ddt2}) we get
\begin{eqnarray}
&&-\frac{1}{4} {\frac {C_1  \left[-2 {\rho_0}^{-\frac{1}{2} C_1  \left( C_1-4 \right) } e^{-C_3} {\dot t}  ^{2}C_2+  {\dot z}  ^{2} \left( C_1-2 \right)  \right] }{ \rho_0}}=0,\\
&&{\ddot z}=0 ,\\
&&{\ddot t}=0 
\end{eqnarray}

Solving these last equations  simultaneously we have
\begin{eqnarray}
t(\tau)&=&C_5 \tau+C_6,\\
z(\tau)&=&\pm {\frac {\sqrt {2}\sqrt {e^{C_3}  \left( C_1-2 \right) {\rho_0}^{-\frac{1}{2} C_1  \left( C_1-4 \right) }C_2} \left( C_5 \tau+C_6 \right) }{e^{C_3}  \left( C_1-2\right) }}+C_4.
\end{eqnarray}

Observe that $C_1>2$ and $C_2>0$  (or $C_1<2$ and $C_2<0$), in order to have real $z$ geodesic.

\begin{figure}[!htp]
	\centering	
	\includegraphics[width=6.5cm]{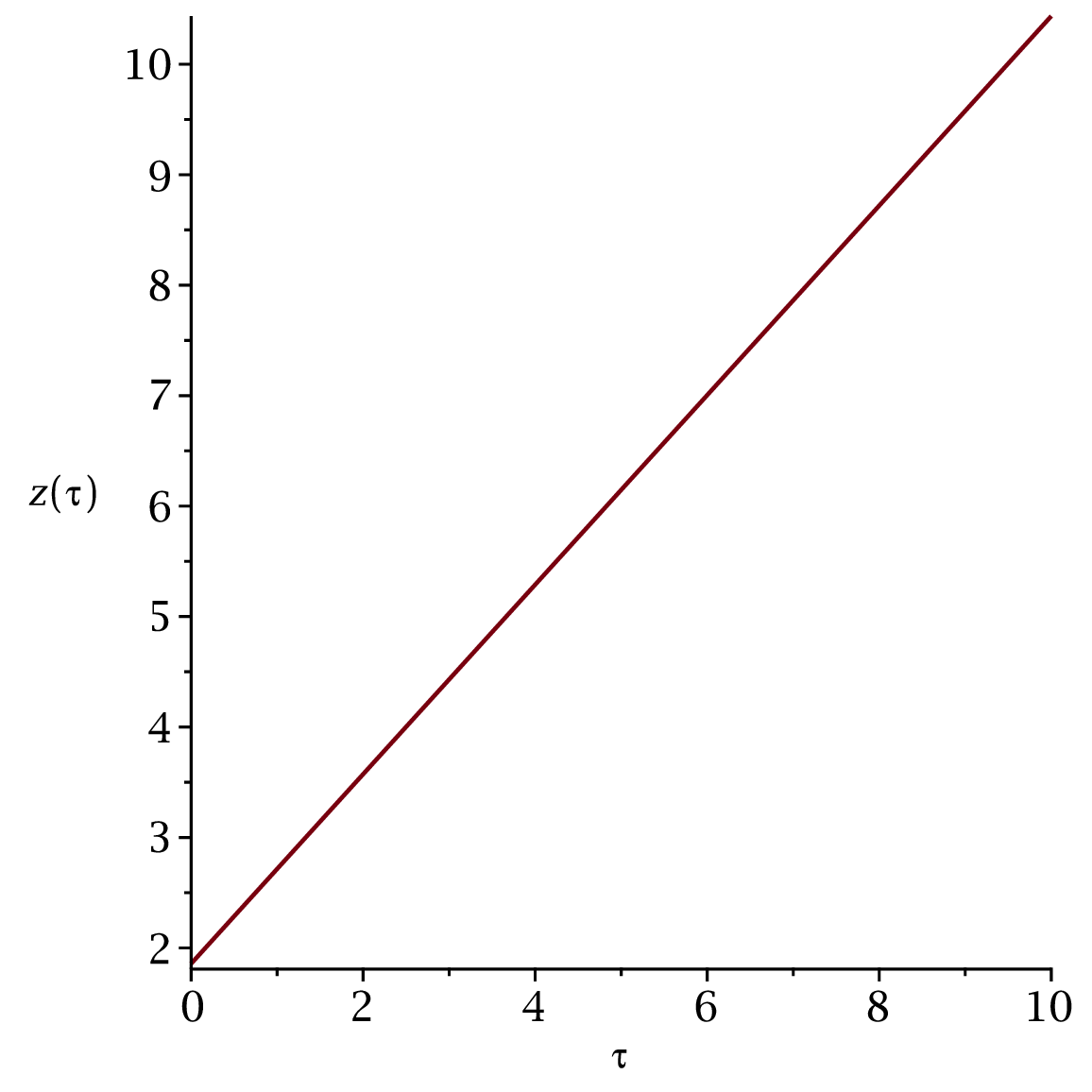}
	\includegraphics[width=6.5cm]{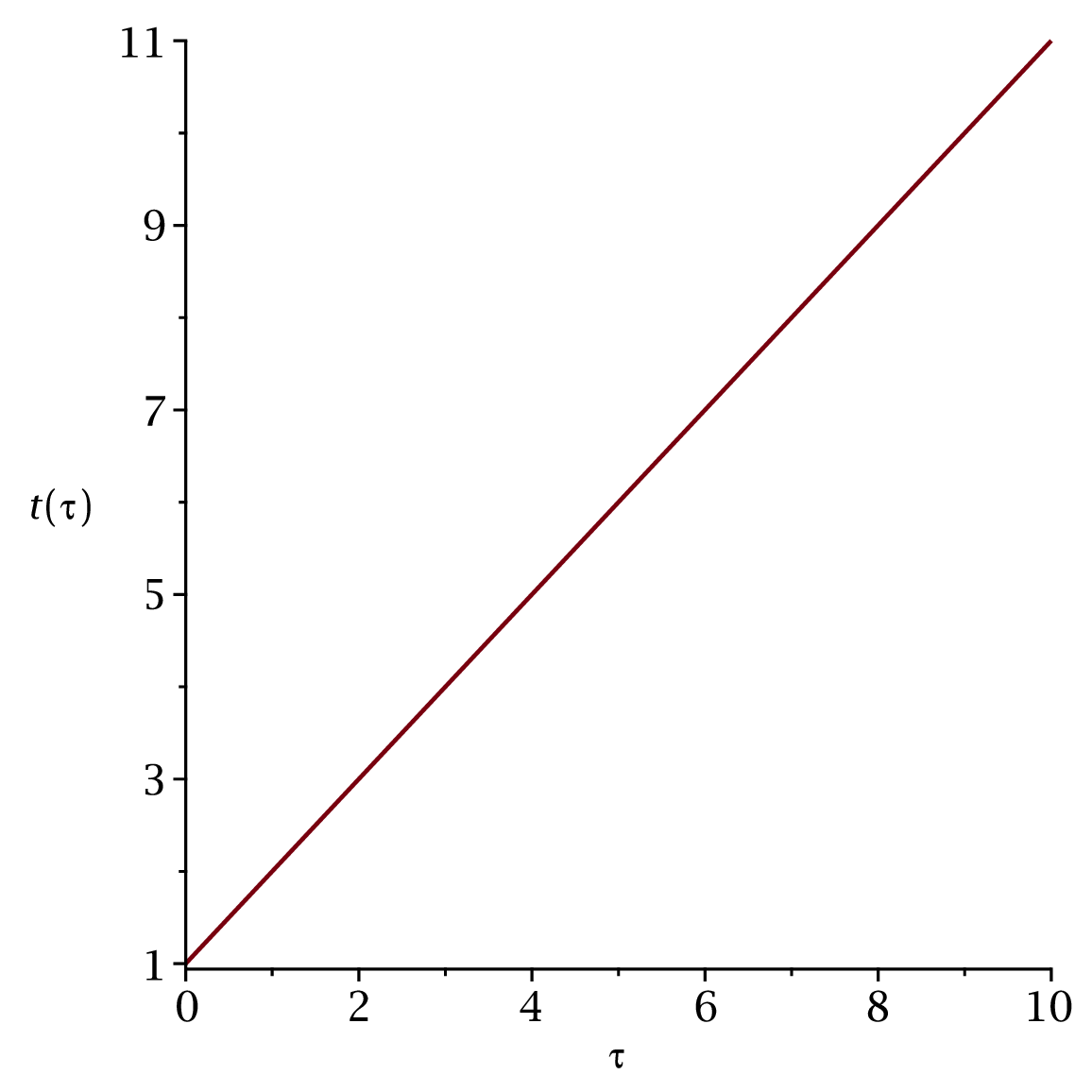}
	\includegraphics[width=6.5cm]{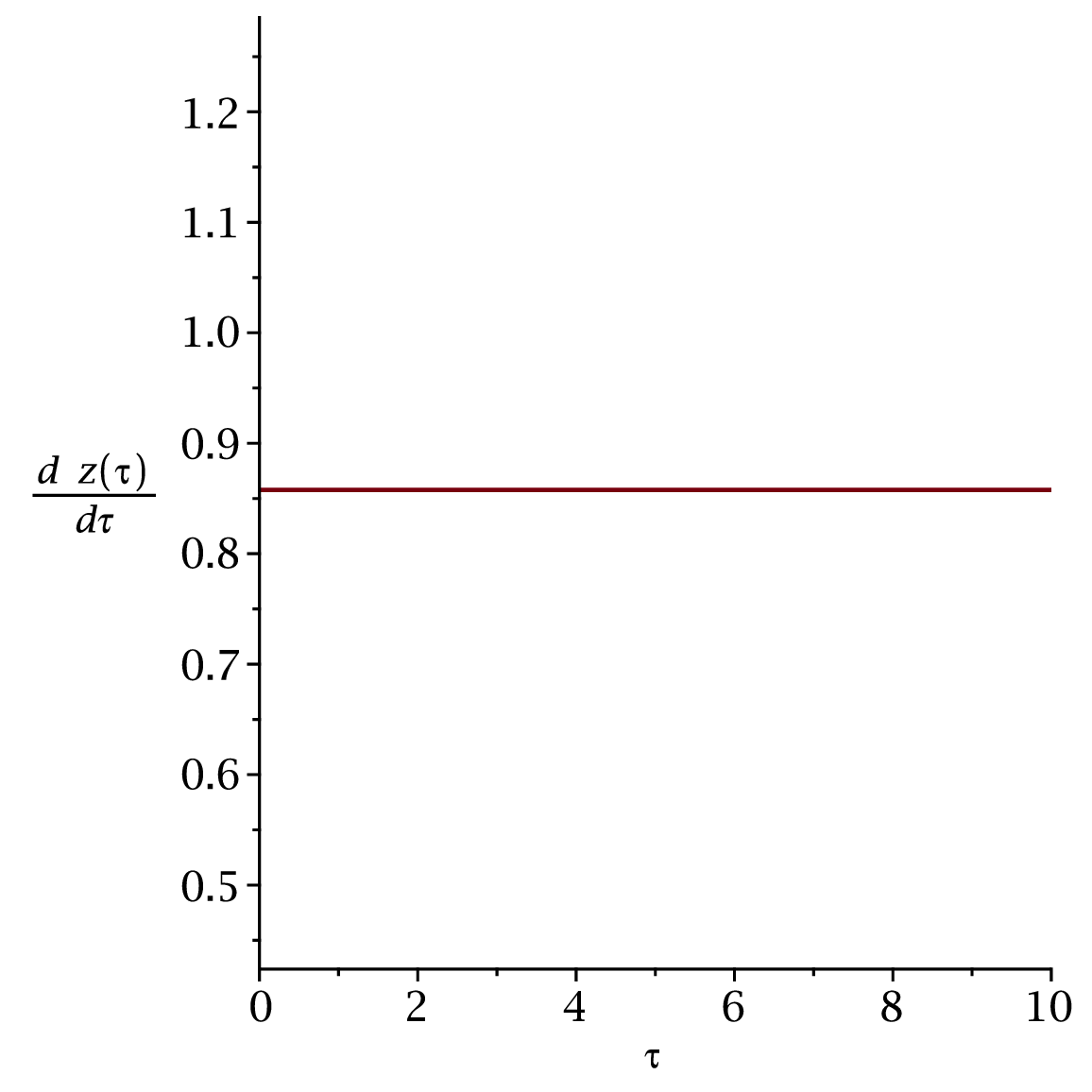}
	\includegraphics[width=6.5cm]{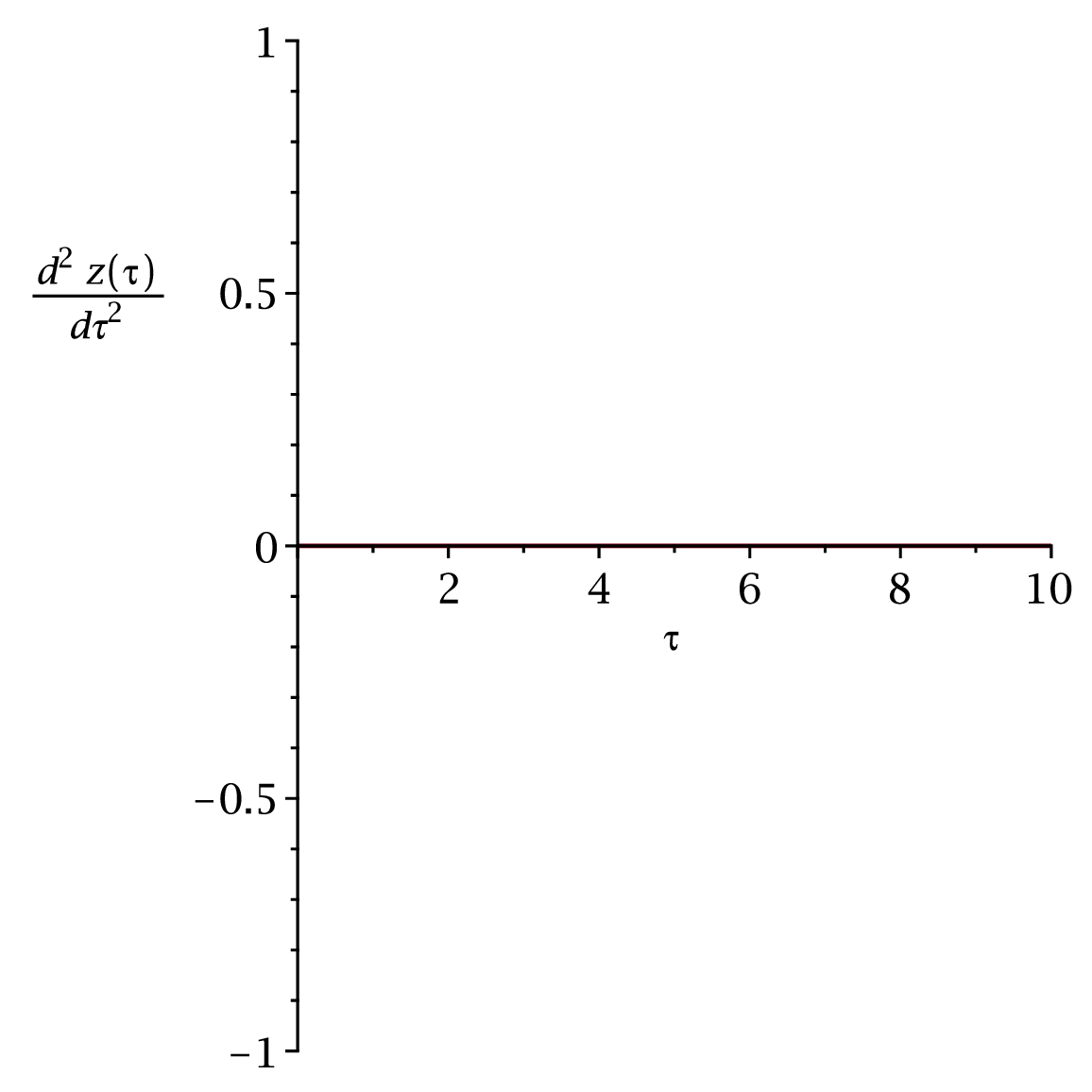}
	\caption{These figures show the time evolution of the $z$-geodesics of the 
	Solution 3 and its velocity and acceleration of $z(\tau)$ at the axis ${\rho_0}=0$.
	We assume in these figures that  $C_1=3$, $C_2=1$, $C_3=1$, $C_4=1$, $C_5=1$
	and $C_6=1$.}
	\label{fig2}
\end{figure}

See Figure \ref{fig2}.

\subsection{Solution 4}

Substituting the  equations (\ref{sol4}) into (\ref{ddr2})-(\ref{ddt2}) we get
\begin{eqnarray}
&&\frac{1}{4}  {\dot z} ^{2}{C_1}^{2}\rho_0=0,\\
&&\ddot z -\frac{1}{2} \dot z ^{2}C_1+\frac{1}{2} {e}^{C_1 z +\frac{1}{4} \rho_0^{2}{C_1}^{2}-C_3}C_1 {e^{C_1 z }}C_2  \dot t ^{2}=0,\\
&&{\ddot t}  +{\dot t}  {\dot z}  C_1=0
\end{eqnarray}

Solving these last equations simultaneously we can  could not obtain an analytical solution.

\subsection{Solution 5(a)}

Substituting the  equations (\ref{sol5a}) into (\ref{ddr2})-(\ref{ddt2}) we get
\begin{eqnarray}
&&{\frac { {\dot t}  ^{2}C_2 \rho_0}{e^{C_1}}}=0,\\
&&{\ddot z}=0,\\
&&{\ddot t}=0. 
\end{eqnarray}

Solving these last equations  simultaneously we have
\begin{eqnarray}
t \left( \tau \right) &=&C_3,\\
z(\tau) &=&C_1 \tau+C_2,
\end{eqnarray}
which is not physically relevant.

\subsection{Solution 5(b)}

Substituting the  equations (\ref{sol5b}) into (\ref{ddr2})-(\ref{ddt2}) we get
\begin{eqnarray}
&&\frac{1}{2} \frac {{e}^{-C_1} \dot t ^{2}{C_3}  \left( 1+\sqrt {1+2 C_2} \right) \rho_0^{-C_2  +\sqrt {1+2 C_2}+1}-C_2  \dot z ^{2}}{\rho_0}=0,\nonumber\\
&&{\ddot z}=0,\nonumber\\
&&{\ddot t}=0. 
\end{eqnarray}

Solving these last equations  simultaneously we have
\begin{eqnarray}
t(\tau)&=&C_5 \tau+C_6,\\
z(\tau)&=&\pm {\frac {\sqrt {e^{C_1} {\rho_0}^{C_2}C_2 C_3 {\rho_0}^{1+\sqrt {1+2 C_2}} \left( 1+\sqrt {1+2 C_2} \right) } \left( C_5 \tau+C_6 \right) }{e^{C_1} {\rho_0}^{C_2}C_2}}+C_4,\nonumber\\
\end{eqnarray}

Observe again that  $C_2 C_3>0$ and $C_2\ge -1/2$ in order to have real $z$ geodesic.

\begin{figure}[!htp]
	\centering	
	\includegraphics[width=6.5cm]{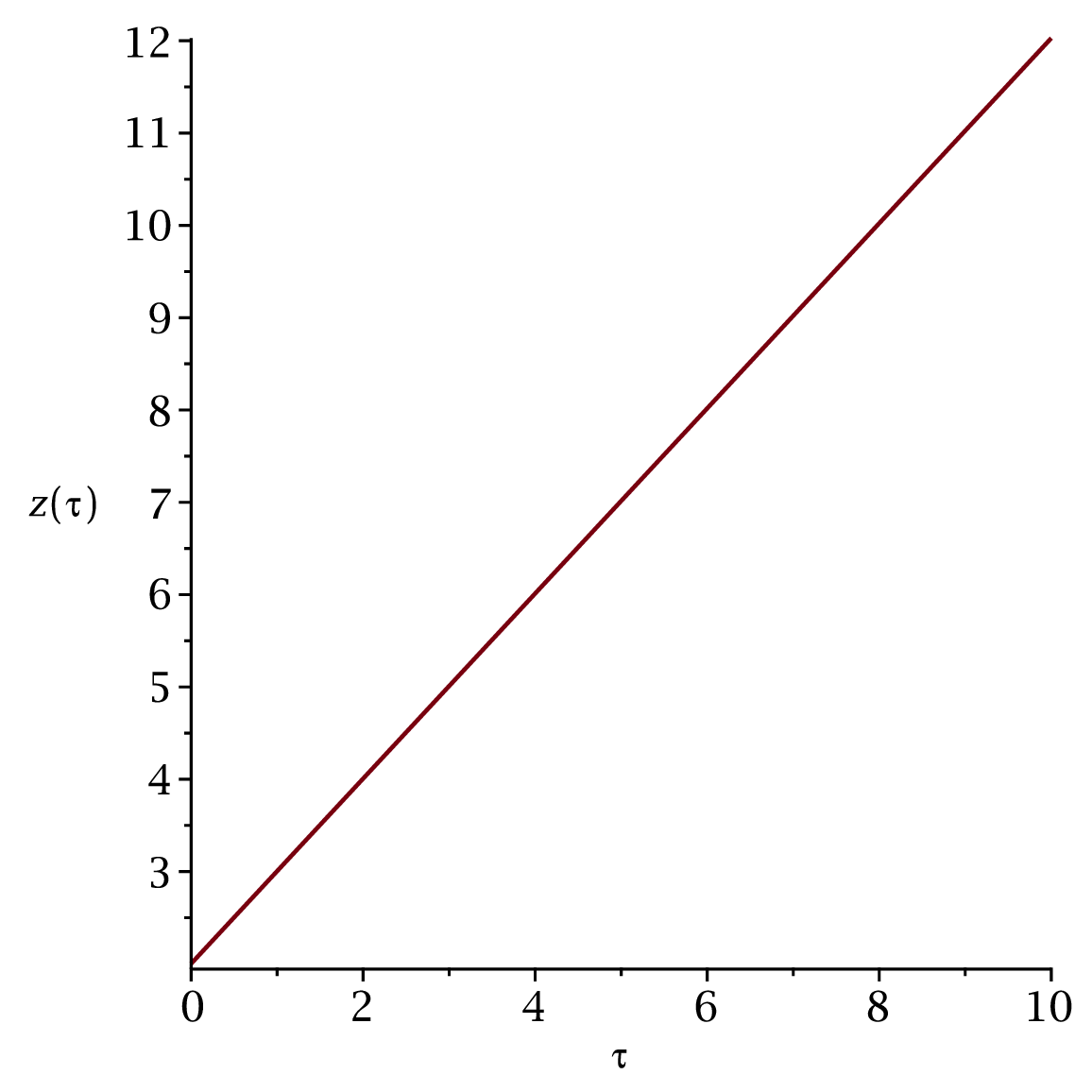}
	\includegraphics[width=6.5cm]{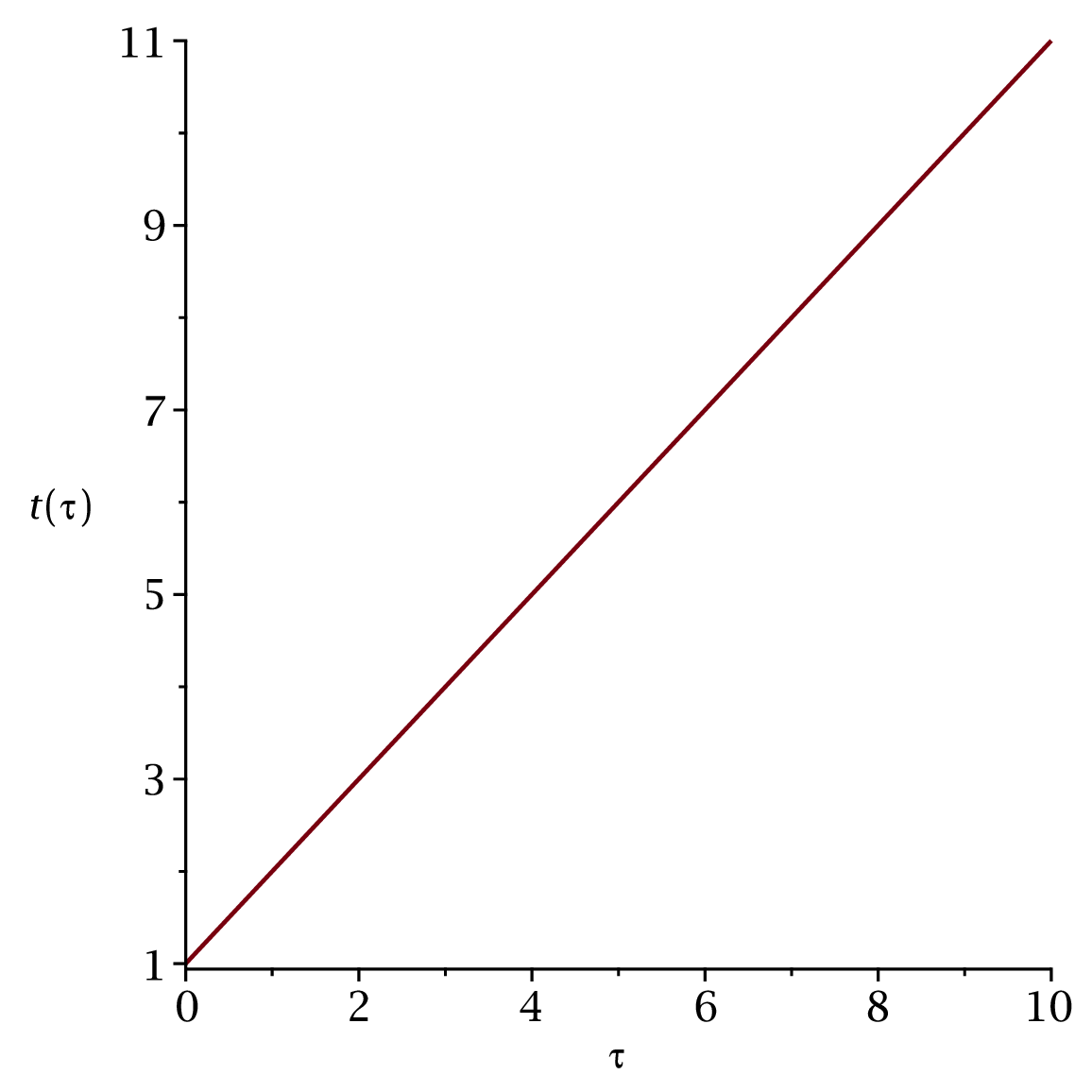}
	\includegraphics[width=6.5cm]{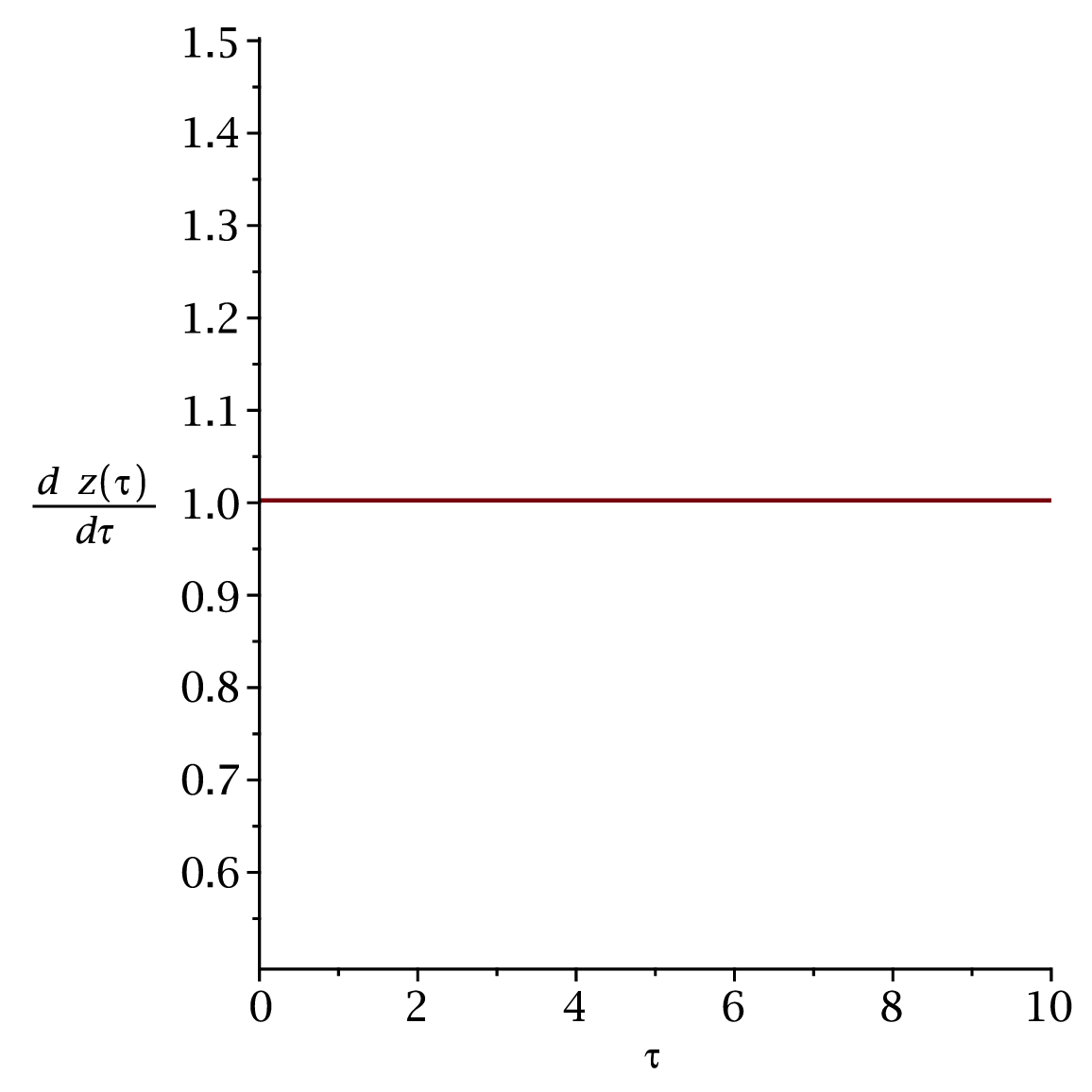}
	\includegraphics[width=6.5cm]{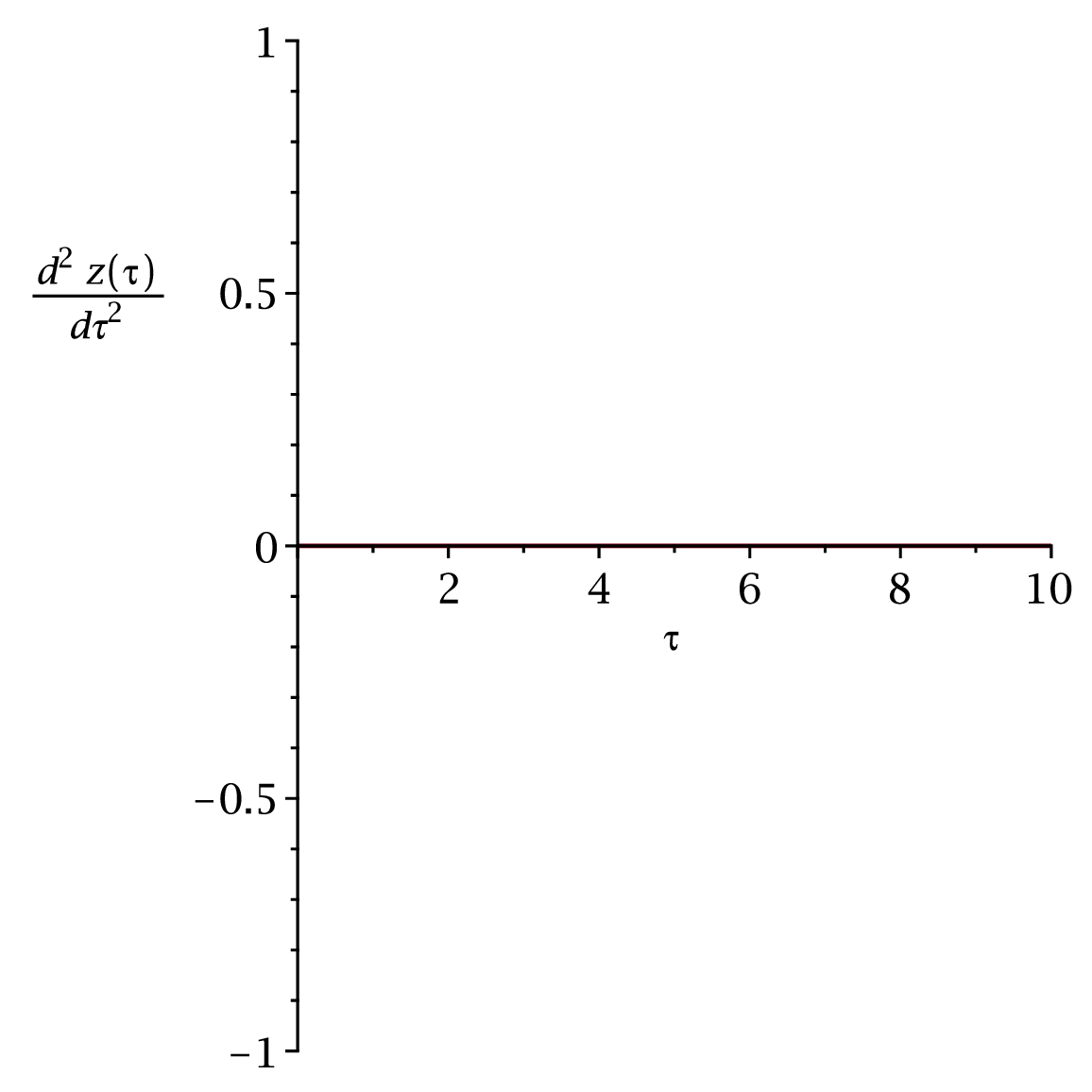}
	\caption{These figures show the time evolution of the $z$-geodesics of the Solution 5b
	and its velocity and acceleration of $z(\tau)$ at the axis ${\rho_0}=1$.
	We assume in these figures that $C_1=1$, $C_2=1$, $C_3=1$, $C_4=1$, $C_5=1$
	and $C_6=1$.}
	\label{fig3}
\end{figure}

See Figure \ref{fig3}.

\subsection{Solution 5(c)}

Substituting the  equations (\ref{sol5c}) into (\ref{ddr2})-(\ref{ddt2}) we get
\begin{eqnarray}
&&-\frac{1}{2\rho_0}  {e^{-\frac{1}{2} {\rho_0}^{2}C_1-C_2-\sqrt {2}\sqrt {-C_1}z }}\times\nonumber\\
&&\left[  \dot z ^{2} \left( {\rho_0}^{2}C_1-\frac{1}{2} \right) {e^{\frac{1}{2} {\rho_0}^{2}C_1+C_2+\sqrt {2}\sqrt {-C_1}z }}-{e^{\sqrt {2}\sqrt {-C_1}C_3}} \dot t ^{2}{\rho_0}^{3/2} \right] =0,\nonumber\\
&&-\frac{1}{2} {e^{-\frac{1}{2} {\rho_0}^{2}C_1-C_2-\sqrt {2}\sqrt {-C_1}z }}  \times\nonumber\\
&&\left(  \dot t ^{2}{e^{\sqrt {2}\sqrt {-C_1}C_3}}{\rho_0}^{3/2}\sqrt {2}\sqrt {-C_1}-2  \ddot z {e^{\frac{1}{2} {\rho_0}^{2}C_1+C_2+\sqrt {2}\sqrt {-C_1}z }} \right)=0.
\end{eqnarray}
Solving these last equations  simultaneously we can note that we cannot obtain an analytical solution.

Thus, for the $\gamma$  Metric and the Solution 2, Solution 4 and Solution 5(c)
 we cannot solve simultaneously the equations (\ref{ddr2})-(\ref{ddt2})
analytically.

\section{Analysis of the Geodesic in z-{\bf $\phi$} Direction}

\subsection{General Solution 1}

Substituting the  equations (\ref{sol1}) into (\ref{ddr3})-(\ref{ddt3}) we get
\begin{eqnarray}
&&-{\frac { {\dot \phi}  ^{2}\rho_0}{ e ^{-\frac{1}{4} {c_1}^{2}\rho_0^{2}+c_3}c_2}}+\frac{1}{4} {\dot z}  ^{2}{c_1}^{2}\rho_0=0,\\
&&{\ddot z}  +\frac{1}{2} {\frac { \left(  e ^{c_1 z } \right) ^{2} {\dot t}  ^{2}c_1 c_2}{ e ^{-\frac{1}{4} {c_1}^{2}\rho_0^{2}+c_3}}}+\frac{1}{2} {\frac { {\dot \phi}  ^{2}\rho_0^{2}c_1}{ e ^{-\frac{1}{4} {c_1}^{2}\rho_0^{2}+c_3}c_2}}-\frac{1}{2}  {\dot z}  ^{2}c_1=0,\\
&&{\ddot z}  -{\dot \phi}  {\dot z}  c_1=0,\\
&&{\ddot t}  +{\dot t}  {\dot z}  c_1=0
\end{eqnarray}

Solving these last equations  simultaneously we can note that we cannot obtain an analytical solution.

\subsection{General Solution 2}

Substituting the  equations (\ref{sol2}) into (\ref{ddr3})-(\ref{ddt3}) we get
\begin{eqnarray}
&&-\frac{1}{4} {\frac { \dot z ^{2}{d_2}^{2}}{\rho_0}}-\frac{1}{2} {\frac { \dot z ^{2}d_2}{\rho_0}}-\frac{1}{16}  \dot z ^{2}{d_0}^{2}{\rho_0}^{3}+\frac{1}{4}  \dot z ^{2}{d_0}^{2}\rho_0 z^{2}+\frac{1}{2}  \dot z ^{2}d_0 d_3 \rho_0z+\nonumber\\
&&\frac{1}{4}  \dot z ^{2}d_0 d_2 \rho_0+\frac{1}{4}  \dot z ^{2}d_0 \rho_0+\frac{1}{4}  \dot z ^{2}\rho_0{d_3}^{2}+\nonumber\\
&&\frac{1}{4}  { \dot t ^{2}d_0 \rho_0{e^{\frac{1}{4} d_0 {\rho_0}^{2}}}{e^{-\frac{1}{2} d_0  z^{2}}}d_1 d_4}\times\nonumber\\
&&\left\{{e}^{\frac{1}{2} \ln  \left( \rho_0 \right) {d_2}^{2}+\ln  \left( \rho_0 \right) d_2+\frac{1}{32} {d_0}^{2}{\rho_0}^{4}-\frac{1}{4} {d_0}^{2}{\rho_0}^{2} z^{2}-\frac{1}{2} d_0 d_3 {\rho_0}^{2}z -\frac{1}{4} d_0 d_2 {\rho_0}^{2}+\frac{1}{2} d_0 d_2  z^{2}}\right. \times\nonumber\\
&&\left. {e}^{-\frac{1}{4} d_0 {\rho_0}^{2}+\frac{1}{2} d_0  z ^{2}-\frac{1}{4} {\rho_0}^{2}{d_3}^{2}+d_3 z d_2+d_3 z +d_5}{e^{d_3 z }}{\rho_0}^{d_2} \right\}^{-1}-\nonumber\\
&&\frac{1}{2}  { \dot t ^{2}{e^{\frac{1}{4} d_0 {\rho_0}^{2}}}{e^{-\frac{1}{2} d_0  z^{2}}}d_1 d_4 d_2} \times\nonumber\\
&&\left\{{e}^{\frac{1}{2} \ln \left( \rho_0 \right) {d_2}^{2}+\ln  \left( \rho_0\right) d_2+\frac{1}{32} {d_0}^{2}{\rho_0}^{4}-\frac{1}{4} {d_0}^{2}{\rho_0}^{2} z^{2}-\frac{1}{2} d_0 d_3 {\rho_0}^{2}z -\frac{1}{4} d_0 d_2 {\rho_0}^{2}+\frac{1}{2} d_0 d_2  z^{2}}\right. \times\nonumber\\
&&\left. {e}^{-\frac{1}{4} d_0 {\rho_0}^{2}+\frac{1}{2} d_0  z^{2}-\frac{1}{4} {\rho_0}^{2}{d_3}^{2}+d_3 z d_2+d_3 z +d_5}{e^{d_3 z }}{\rho_0}^{d_2}\rho_0\right\}^{-1} \nonumber\\
&&- { \dot \phi ^{2}\rho_0{e^{d_3 z }}{\rho_0}^{d_2}} \left\{{e}^{\frac{1}{2} \ln  \left( \rho_0\right) {d_2}^{2}+\ln  \left( \rho_0 \right) d_2+\frac{1}{32} {d_0}^{2}{\rho_0}^{4}-\frac{1}{4} {d_0}^{2}{\rho_0}^{2} z^{2}-\frac{1}{2} d_0 d_3 {\rho_0}^{2}z }\right.\times\nonumber\\
&&\left. {e}^{-\frac{1}{4} d_0 d_2 {\rho_0}^{2}+\frac{1}{2} d_0 d_2  z^{2}-\frac{1}{4} d_0 {\rho_0}^{2}+\frac{1}{2} d_0  z^{2}-\frac{1}{4} {\rho_0}^{2}{d_3}^{2}+d_3 z d_2+d_3 z +d_5}{e^{\frac{1}{4} d_0 {\rho_0}^{2}}}{e^{-\frac{1}{2} d_0  z^{2}}}d_1 d_4 \right\}^{-1} +\nonumber\\
&&\frac{1}{4}  { \dot \phi ^{2}{\rho_0}^{3}{e^{d_3 z }}{\rho_0}^{d_2}d_0} \left\{{e}^{\frac{1}{2} \ln  \left( \rho_0 \right) {d_2}^{2}+\ln  \left( \rho_0 \right) d_2+\frac{1}{32} {d_0}^{2}{\rho_0}^{4}-\frac{1}{4} {d_0}^{2}{\rho_0}^{2} z^{2}-\frac{1}{2} d_0 d_3 {\rho_0}^{2}z }\right.\times\nonumber\\
&&\left. {e}^{-\frac{1}{4} d_0 d_2 {\rho_0}^{2}+\frac{1}{2} d_0 d_2  z^{2}-\frac{1}{4} d_0 {\rho_0}^{2}+\frac{1}{2} d_0  z ^{2}-\frac{1}{4} {\rho_0}^{2}{d_3}^{2}+d_3 z d_2+d_3 z +d_5}{e^{\frac{1}{4} d_0 {\rho_0}^{2}}}{e^{-\frac{1}{2} d_0  z ^{2}}}d_1 d_4\right\}^{-1}-\nonumber\\
&&\frac{1}{2}  { \dot \phi ^{2}\rho_0{e^{d_3 z }}{\rho_0}^{d_2}d_2} \left\{{e}^{\frac{1}{2} \ln  \left( \rho_0 \right) {d_2}^{2}+\ln \left( \rho_0 \right) d_2+\frac{1}{32} {d_0}^{2}{\rho_0}^{4}-\frac{1}{4} {d_0}^{2}{\rho_0}^{2} z ^{2}-\frac{1}{2} d_0 d_3 {\rho_0}^{2}z -\frac{1}{4} d_0 d_2 {\rho_0}^{2}}\right. \times\nonumber\\
&&\left. {e}^{\frac{1}{2} d_0 d_2  z^{2}-\frac{1}{4} d_0 {\rho_0}^{2}+\frac{1}{2} d_0  z^{2}-\frac{1}{4} {\rho_0}^{2}{d_3}^{2}+d_3 z d_2+d_3 z +d_5}{e^{\frac{1}{4} d_0 {\rho_0}^{2}}}{e^{-\frac{1}{2} d_0  z^{2}}}d_1 d_4 \right\}^{-1}=0,\\
&&\ddot z -\frac{1}{4} \dot z ^{2}{d_0}^{2}{\rho_0}^{2}z -\frac{1}{4}  \dot z ^{2}d_0 d_3 {\rho_0}^{2}+\frac{1}{2}  \dot z ^{2}d_0 d_2 z +\frac{1}{2}  \dot z ^{2}d_0 z +\frac{1}{2}  \dot z ^{2}d_3 d_2+\frac{1}{2}  \dot z ^{2}d_3-\nonumber\\
&&\frac{1}{2}  { \dot t ^{2}{e^{\frac{1}{4} d_0 {\rho_0}^{2}}}d_0 z {e^{-\frac{1}{2} d_0  z^{2}}}d_1 d_4} \left\{{e}^{\frac{1}{2} \ln  \left( \rho_0 \right) {d_2}^{2}+\ln \left( \rho_0 \right) d_2+\frac{1}{32} {d_0}^{2}{\rho_0}^{4}-\frac{1}{4} {d_0}^{2}{\rho_0}^{2} z ^{2}} \right.\times\nonumber\\
&&\left. {e}^{-\frac{1}{2} d_0 d_3 {\rho_0}^{2}z -\frac{1}{4} d_0 d_2 {\rho_0}^{2}+\frac{1}{2} d_0 d_2  z^{2}-\frac{1}{4} d_0 {\rho_0}^{2}+\frac{1}{2} d_0  z^{2}-\frac{1}{4} {\rho_0}^{2}{d_3}^{2}+d_3 z d_2+d_3 z +d_5}{e^{d_3 z }}{\rho_{{0}}}^{d_2}\right\}^{-1}-\nonumber\\
&&\frac{1}{2}  { \dot t ^{2}{e^{\frac{1}{4} d_0 {\rho_0}^{2}}}{e^{-\frac{1}{2} d_0  z^{2}}}d_1 d_4 d_3} \times\nonumber\\
&&\left\{{e}^{\frac{1}{2} \ln  \left( \rho_0\right) {d_2}^{2}+\ln  \left( \rho_0 \right) d_2+\frac{1}{32} {d_0}^{2}{\rho_0}^{4}-\frac{1}{4} {d_0}^{2}{\rho_0}^{2} z^{2}-\frac{1}{2} d_0 d_3 {\rho_0}^{2}z -\frac{1}{4} d_0 d_2 {\rho_0}^{2}+\frac{1}{2} d_0 d_2  z^{2}}\right.\times\nonumber\\
&&\left. {e}^{-\frac{1}{4} d_0 {\rho_0}^{2}+\frac{1}{2} d_0  z^{2}-\frac{1}{4} {\rho_0}^{2}{d_3}^{2}+d_3 z d_2+d_3 z +d_5}{e^{d_3 z }}{\rho_0}^{d_2} \right\}^{-1}-\nonumber\\
&&\frac{1}{2}  {\dot \phi ^{2}{\rho_0}^{2}{e^{d_3 z }}{\rho_0}^{d_2}d_0 z }  \left\{{e}^{\frac{1}{2} \ln  \left( \rho_0 \right) {d_2}^{2}+\ln  \left( \rho_0 \right) d_2+\frac{1}{32} {d_0}^{2}{\rho_0}^{4}-\frac{1}{4} {d_0}^{2}{\rho_0}^{2} z^{2}-\frac{1}{2} d_0 d_3 {\rho_0}^{2}z -\frac{1}{4} d_0 d_2 {\rho_0}^{2}}\right.\times\nonumber\\
&&\left. {e}^{\frac{1}{2} d_0 d_2  z^{2}-\frac{1}{4} d_0 {\rho_0}^{2}+\frac{1}{2} d_0  z^{2}-\frac{1}{4} {\rho_0}^{2}{d_3}^{2}+d_3 z d_2+d_3 z +d_5}{e^{\frac{1}{4} d_0 {\rho_0}^{2}}}{e^{-\frac{1}{2} d_0  z^{2}}}d_1 d_4 \right\}^{-1}-\nonumber\\
&&\frac{1}{2}  { \dot \phi ^{2}{\rho_0}^{2}d_3 {e^{d_3 z }}{\rho_0}^{d_2}} \left\{{e}^{\frac{1}{2} \ln  \left( \rho_0 \right) {d_2}^{2}+\ln  \left( \rho_0 \right) d_2+\frac{1}{32} {d_0}^{2}{\rho_0}^{4}-\frac{1}{4} {d_0}^{2}{\rho_0}^{2} z^{2}}\right.\times\nonumber\\
&&\left.{e}^{-\frac{1}{2} d_0 d_3 {\rho_0}^{2}z -\frac{1}{4} d_0 d_2 {\rho_0}^{2}+\frac{1}{2} d_0 d_2  z^{2}-\frac{1}{4} d_0 {\rho_0}^{2}+\frac{1}{2} d_0  z ^{2}-\frac{1}{4} {\rho_0}^{2}{d_3}^{2}+d_3 z d_2+d_3 z }\right.\times\nonumber\\
&&\left. {e}^{d_5}{e^{\frac{1}{4} d_0 {\rho_0}^{2}}}{e^{-\frac{1}{2} d_0  z ^{2}}}d_1 d_4\right\}=0,\\
&&{\ddot z}=0,\\
&&{\ddot \phi}=0,\\
&&{\ddot t}=0  
\end{eqnarray}

Solving these last equations simultaneously we can note that we cannot obtain an analytical solution.

\subsection{Solution 3}

Substituting the  equations (\ref{sol3}) into (\ref{ddr3})-(\ref{ddt3}) we get

\begin{eqnarray}
t \left( \tau \right) &=&C_7 \tau+C_8,\\
z(\tau) &=&C_5 \tau+C_6,\\
\phi(\tau)&=&\pm \frac{1}{2{ \left( C_1-2 \right) \rho_0}} \sqrt {2}
\left[ {C_1}^{3}C_2 \rho_0^{\frac{1}{2} {C_1}^{2}}{C_5}^{2}C_3-4 {C_1}^{2}C_2 \rho_0^{\frac{1}{2} {C_1}^{2}}{C_5}^{2}C_3- \right.\nonumber\\
&&\left. 2 {C_1}^{2}{C_2}^{2} \left( \rho_0^{C_1}\right) ^{2}{C_7}^{2}+4 C_1 C_2 \rho_0^{\frac{1}{2} {C_1}^{2}}{C_5}^{2}C_3+\right.\nonumber\\
&&\left. 4 C_1 {C_2}^{2} \left( \rho_0^{C_1} \right) ^{2}{C_7}^{2} \right]^{\frac{1}{2}}\tau+C_4\nonumber\\
\end{eqnarray}
 describing an helical motion of a test particle along the $z$ axis.

\begin{figure}[!htp]
	\centering
	\includegraphics[width=6.5cm]{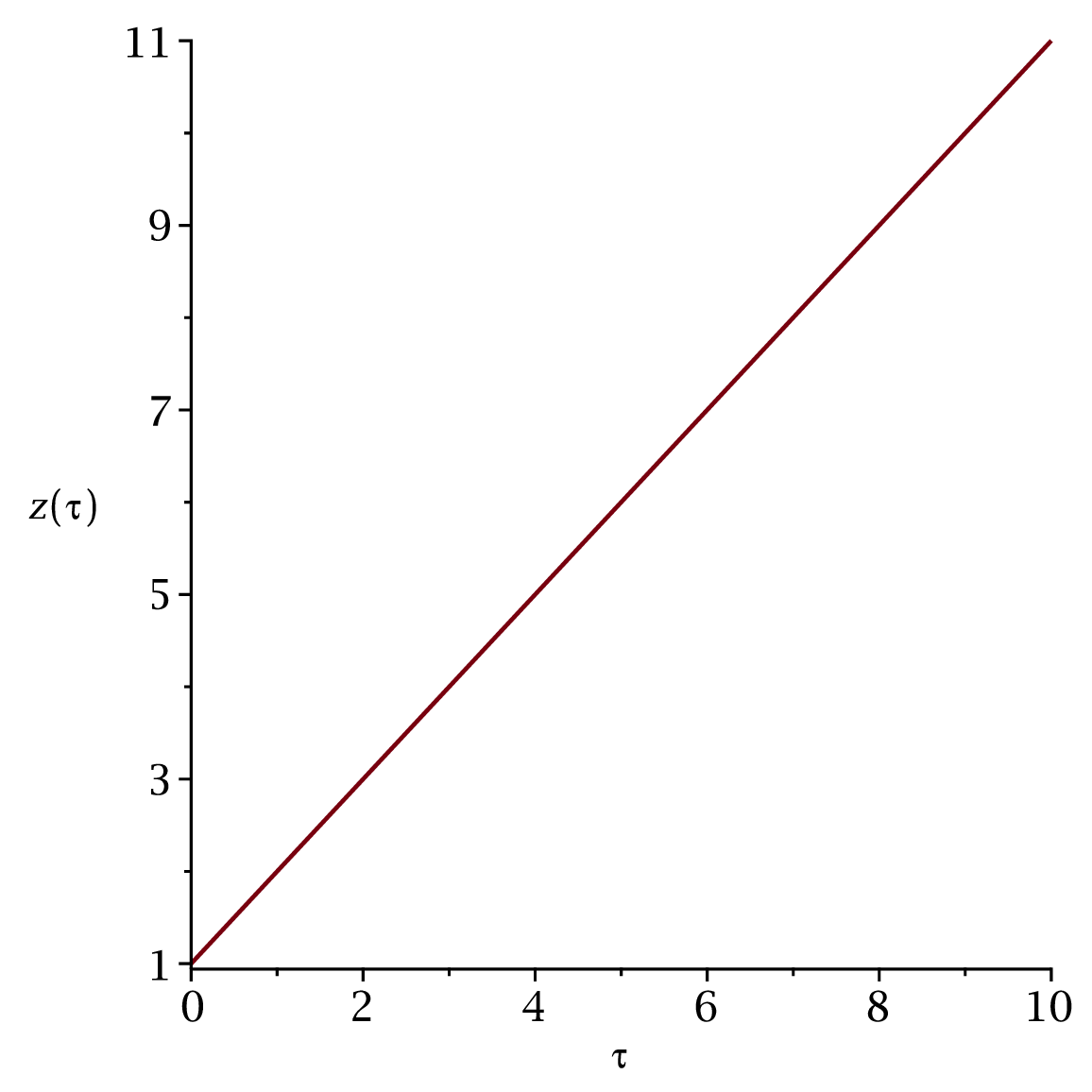}
	\includegraphics[width=6.5cm]{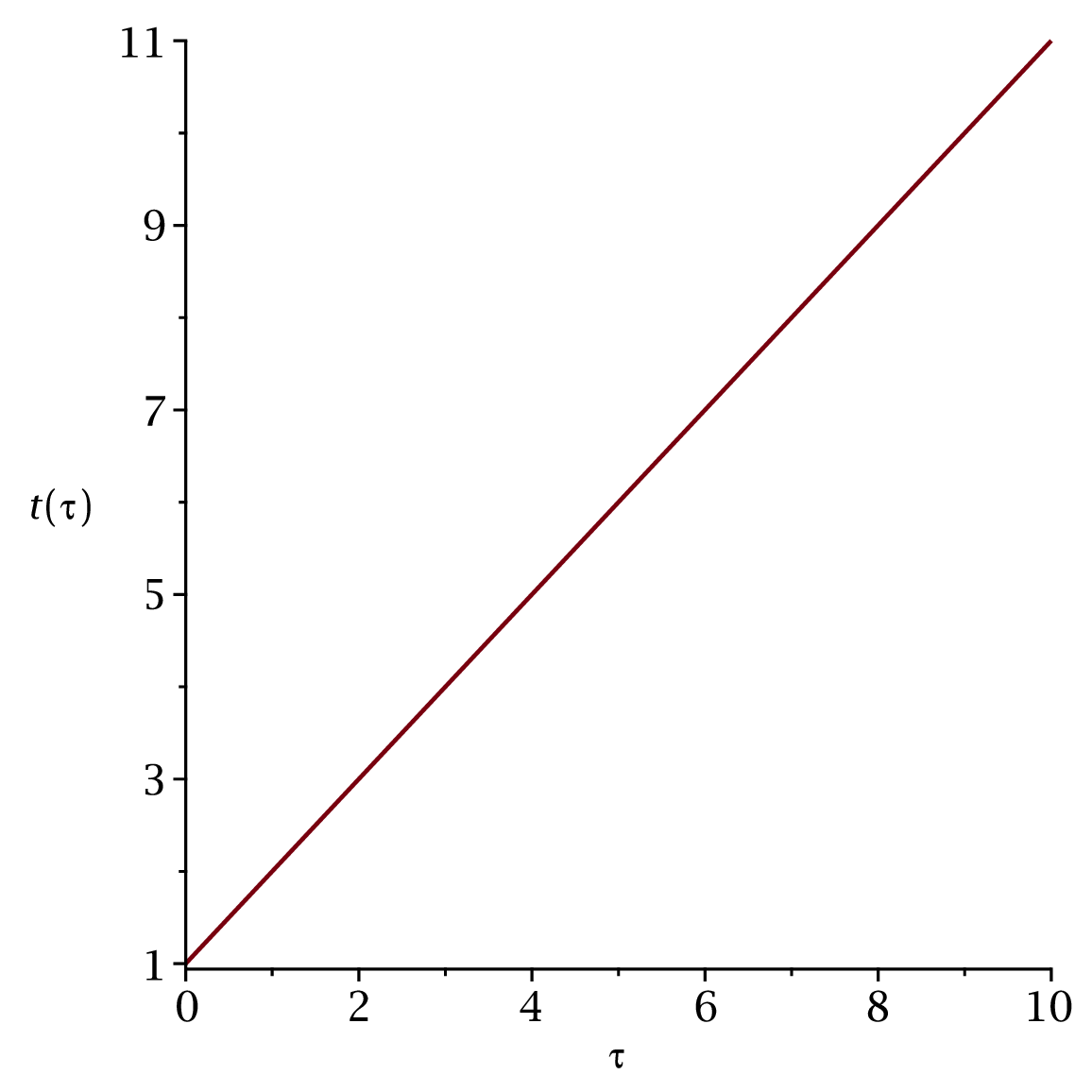}
	\includegraphics[width=8cm]{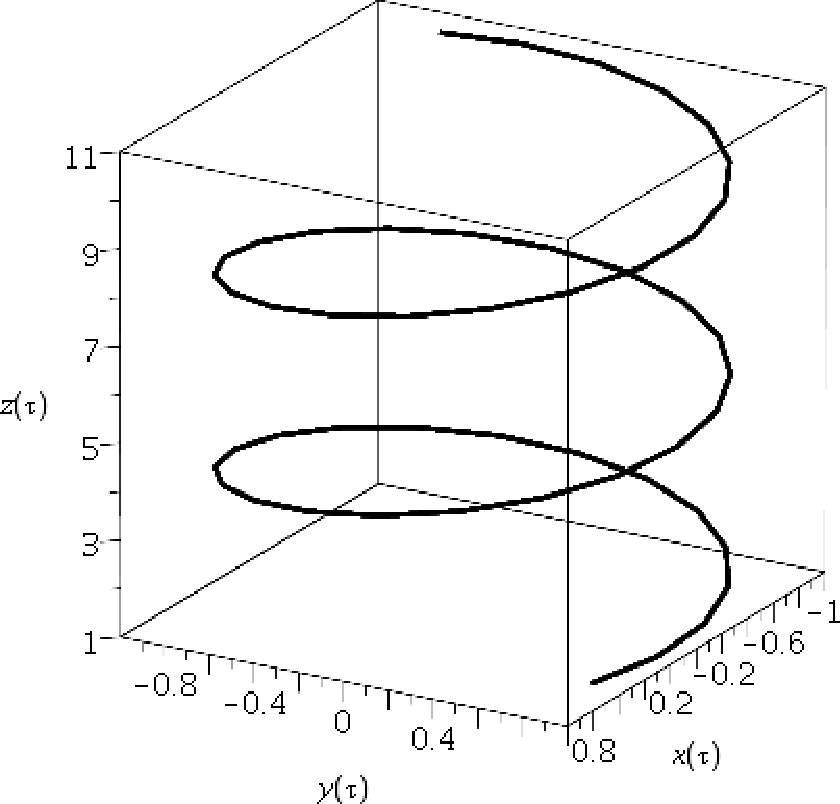}
	\caption{These figures show the time evolution of the $z-\phi$ geodesics of the 
		Solution 3 at the cylinder ${\rho_0}=1$.
		We assume in these figures that $C_1=1$, $C_2=-1$, $C_3=1$, $C_4=1$, $C_5=1$,
		$C_6=1$, $C_7=1/10$ and $C_8=1$.}
	\label{fig5}
\end{figure}

See Figures \ref{fig5}.

\subsection{Solution 4}

Substituting the  equations (\ref{sol4}) into (\ref{ddr3})-(\ref{ddt3}) we get
\begin{eqnarray}
&&-{\frac { {\dot \phi} ^{2}\rho_0}{C_3 {e^{-\frac{1}{4} {C_1}^{2}\rho_0^{2}}}C_2}}+\frac{1}{4}  {\dot z}  ^{2}{C_1}^{2}\rho_0=0,\\
&&{\ddot z}  +\frac{1}{2} {\frac { \left( {e^{z C_1}} \right) ^{2} {\dot t}  ^{2}C_1 C_2}{C_3 {e^{-\frac{1}{4} {C_1}^{2}\rho_0^{2}}}}}+\frac{1}{2} {\frac { {\dot \phi}  ^{2}\rho_0^{2}C_1}{C_3 {e^{-\frac{1}{4} {C_1}^{2}\rho_0^{2}}}C_2}}-\frac{1}{2}  {\dot z}  ^{2}C_1=0,\\
&&{\ddot \phi}  -{\dot \phi}  {\dot z}  C_1=0,\\
&&{\ddot t}  +{\dot t}  {\dot z}  C_1=0
\end{eqnarray}

Solving these last equations  simultaneously we can note that we cannot obtain an analytical solution.

\subsection{Solution 5(a)}

Substituting the  equations (\ref{sol5a}) into (\ref{ddr3})-(\ref{ddt3}) we get
\begin{eqnarray}
&&{\frac { {\dot t}  ^{2}C_2 \rho_0}{C_1}}=0,\\
&&{\ddot z}=0,\\
&&{\ddot \phi}=0,\\
&&{\ddot t}=0 
\end{eqnarray}

Solving these last equations  simultaneously we have
\begin{eqnarray}
t(\tau)&=&C_5,\\
z(\tau)&=&C_3 \tau+C_4,\\
\phi(\tau)&=&C_1 \tau+C_2,
\end{eqnarray}
thus, we have a solution not relevant physically.

\subsection{Solution 5(b)}

Substituting the  equations (\ref{sol5b}) into (\ref{ddr3})-(\ref{ddt3}) we get
\begin{eqnarray}
&&\frac{1}{2} {\frac { {\dot t}  ^{2}C_3 \rho_0^{\sqrt {1+2 C_1}}}{\rho_0^{C_1}C_2}}+\frac{1}{2} {\frac { {\dot t}  ^{2}C_3 \rho_0^{\sqrt {1+2 C_1}}\sqrt {1+2 C_1}}{\rho_0^{C_1}C_2}}-\frac{1}{2} {\frac { {\dot \phi}  ^{2}}{\rho_0^{C_1}C_2 C_3 \rho_0^{\sqrt {1+2 C_1}}}}+\nonumber\\
&&\frac{1}{2} {\frac {{\dot \phi}  ^{2}\sqrt {1+2 C_1}}{\rho_0^{C_1}C_2 C_3 \rho_0^{\sqrt {1+2 C_1}}}}-\frac{1}{2} {\frac { {\dot z}  ^{2}C_1}\rho_0}=0,\\
&&{\ddot z}=0,\\
&&{\ddot \phi}=0,\\
&&{\ddot t}=0  
\end{eqnarray}

Solving these last equations  simultaneously we get
\begin{eqnarray}
t \left( \tau \right) &=&C_7 \tau+C_8,\\
z(\tau) &=&C_5 \tau+C_6,\\
\phi(\tau)&=&\frac {1}{{\rho_{{0}} \left( \sqrt {1+2 C_1}-1 \right) }}\times\nonumber\\
&&\left[ C_3 \rho_{{0}}\sqrt {1+2 C_1}{\rho_{{0}}}^{C_1}{\rho_{{0}}}^{\sqrt {1+2 C_1}}{C_5}^{2}C_1 C_2- \right.\nonumber\\
&&\left. 2 {C_3}^{2}{\rho_{{0}}}^{2} \left( {\rho_{{0}}}^{\sqrt {1+2 C_1}} \right) ^{2}{C_7}^{2}C_1-C_3 \rho_{{0}}{\rho_{{0}}}^{C_1}{\rho_{{0}}}^{\sqrt {1+2 C_1}}{C_5}^{2}C_1 C_2 \right]^{\frac{1}{2}}\tau+C_4.\nonumber\\
\end{eqnarray}

Note that, in principle, we must have $C_1>-1/2$ in order to have real geodesic.
 The motion of the test particle is similar to that of the Solution 3.

\begin{figure}[!htp]
	\centering
	\includegraphics[width=6.5cm]{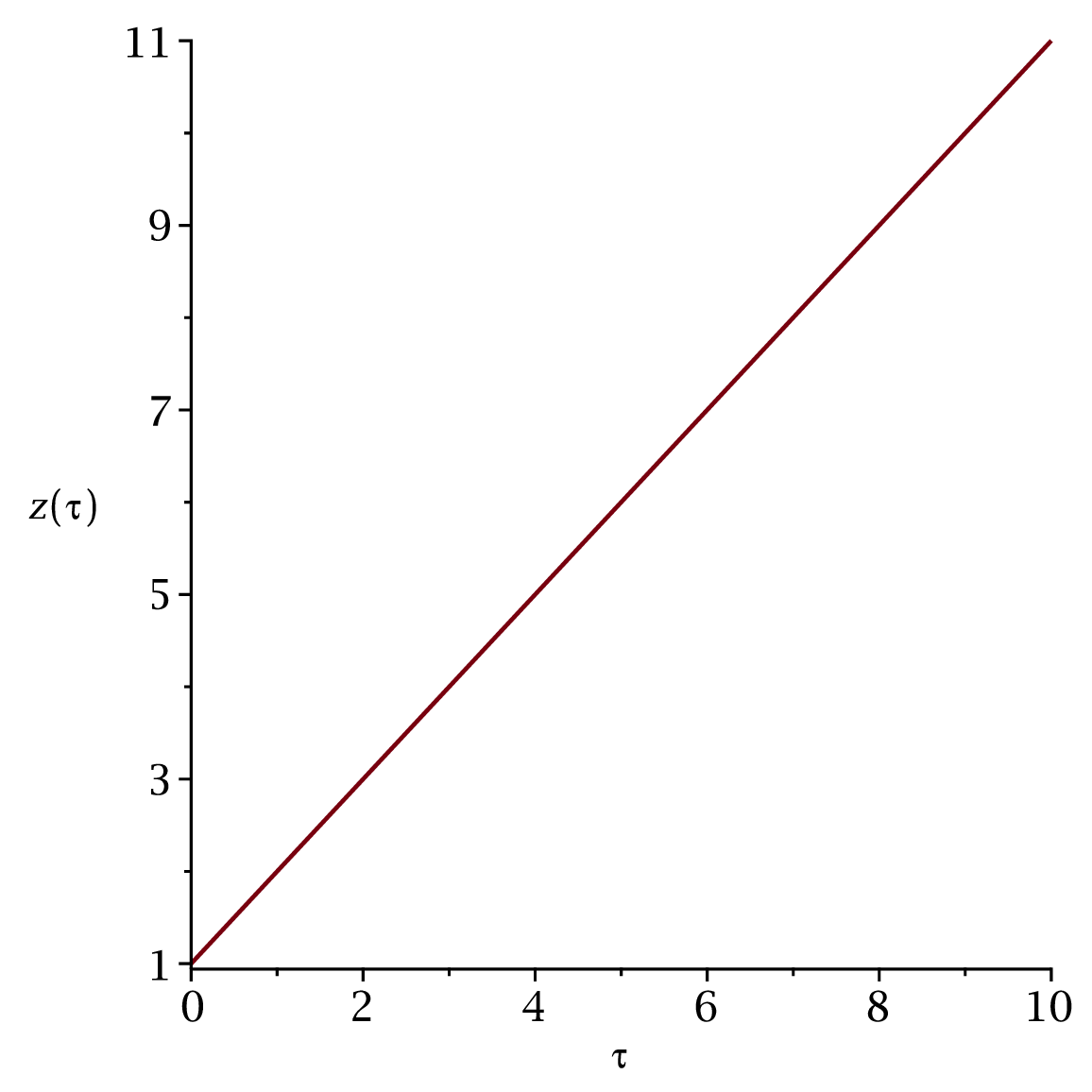}
	\includegraphics[width=6.5cm]{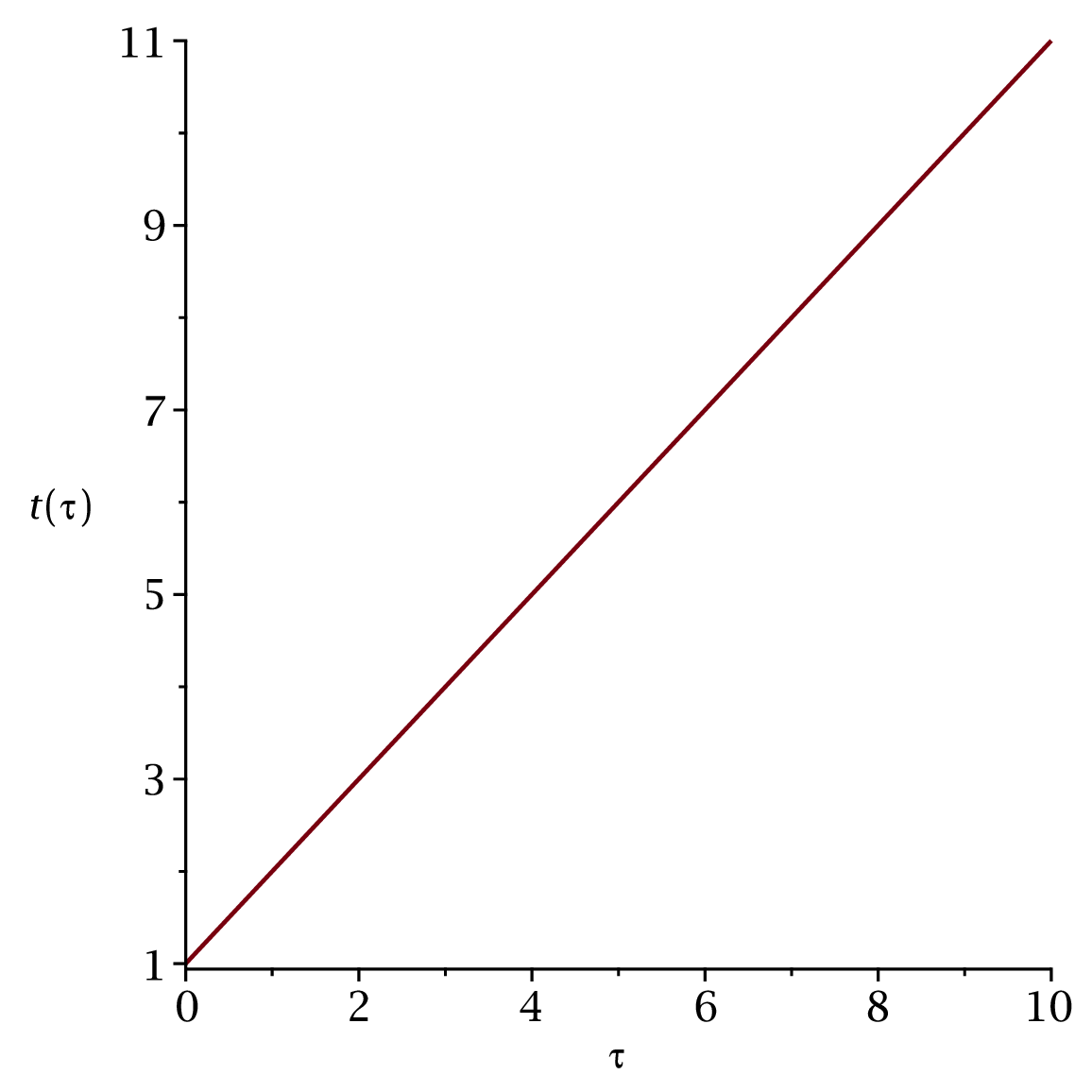}
	\includegraphics[width=8cm]{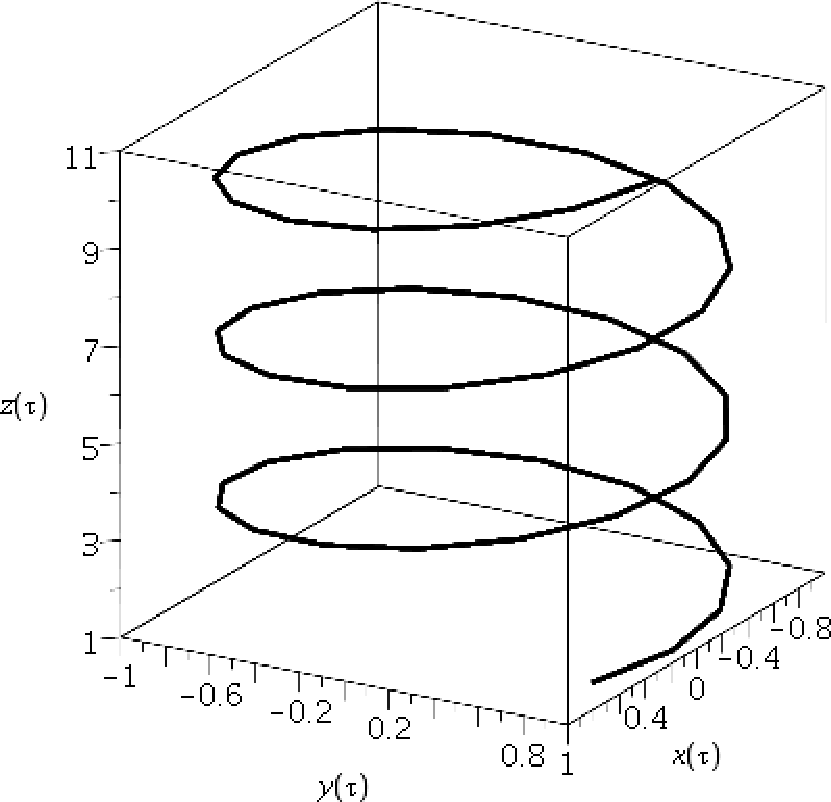}
	\caption{These figures show the time evolution of the $z-\phi$ geodesics of the 
	Solution 5b at the cylinder ${\rho_0}=1$.
	We assume in these figures that $C_1=1$, $C_2=1$, $C_3=1$, $C_4=1$, $C_5=1$,
	$C_6=1$, $C_7=1$ and $C_8=1$.}
	\label{fig4}
\end{figure}

See Figures \ref{fig4}.

\subsection{Solution 5(c)}

Substituting the  equations (\ref{sol5c}) into (\ref{ddr3})-(\ref{ddt3}) we get
\begin{eqnarray}
&&\frac{1}{2} {\frac {\sqrt \rho_0 {\dot t}  ^{2}{e^{\sqrt {C_1}\sqrt {2}C_3}}}{{e^{-\frac{1}{2} C_1 \rho_0^{2}}}C_2 {e^{-\frac{1}{4}}}{e^{\sqrt {C_1}\sqrt {2}z }}}}-\frac{1}{2} {\frac {\sqrt \rho_0{e^{\sqrt {C_1}\sqrt {2}z }} {\dot \phi}  ^{2}}{{e^{-\frac{1}{2} C_1 \rho_0^{2}}}C_2 {e^{-\frac{1}{4}}}{e^{\sqrt {C_1}\sqrt {2}C_3}}}}+\frac{1}{2} \rho_0 {\dot z}  ^{2}C_1+\frac{1}{4} {\frac { {\dot z}  ^{2}}\rho_0}=0,\nonumber\\
\\
&&{\ddot z}  -\frac{1}{2} {\frac {\rho_0^{\frac{3}{2}}{e^{\sqrt {C_1}\sqrt {2}C_3}}\sqrt {C_1}\sqrt {2} {\dot t}  ^{2}}{{e^{-\frac{1}{2} C_1 \rho_0^{2}}}C_2 {e^{-\frac{1}{4}}}{e^{\sqrt {C_1}\sqrt {2}z }}}}-\frac{1}{2} {\frac {\rho_0^{\frac{3}{2}}\sqrt {C_1}\sqrt {2}{e^{\sqrt {C_1}\sqrt {2}z }} {\dot \phi}  ^{2}}{{e^{-\frac{1}{2} C_1 \rho_0^{2}}}C_2 {e^{-\frac{1}{4}}}{e^{\sqrt {C_1}\sqrt {2}C_3}}}}=0,\\
&&{\ddot \phi}  +{\dot \phi}  {\dot z}  \sqrt {C_1}\sqrt {2}{\ddot t}  -{\dot t}  {\dot z}  \sqrt {C_1}\sqrt {2}=0,\\
&& {\ddot t}  -{\dot t}  {\dot z}  \sqrt {C_1}\sqrt {2}=0
\end{eqnarray}

Solving these last equations simultaneously we can note that we cannot obtain an analytical solution.

\section{Analysis of the Geodesic in {\bf $\rho$}-Direction}
For the sake of simplicity we present only the radial geodesic of the Solution 5(a)
because we cannot solve simultaneously the equations (\ref{ddr4})-(\ref{ddt4})
analytically for the others solutions.

\subsection{Solution 5(a)}

Substituting the  equations (\ref{sol5a}) into (\ref{ddr4})-(\ref{ddt4}) we get
\begin{eqnarray}
&&{\ddot \rho}  +{\frac 
{ {\dot t}  ^{2}C_2 \rho }{e^{C_1}}}=0,\nonumber\\
&&{\ddot t}  +2 {\frac { {\dot t}  {\dot \rho}  }{\rho }}=0.\label{geor}
\end{eqnarray}
For the Solution 1, assuming that ${z}_0 \neq 0$, we have that
solving simultaneously the equations (\ref{geor}) we get
\begin{eqnarray}
t(\tau)&=&{\frac {e^{C_1}}{\sqrt {e^{C_1} C_2}}{\rm arctanh} \left(-\frac{1}{4} {\frac {{C_4}^{2}C_5+{C_4}^{2}\tau}{C_4 \sqrt {e^{C_1} C_2}}}\right)}+C_6,\\
\rho(t)&=&\frac{1}{2} {C_3{\frac {1}{\sqrt {-{\frac {e^{C_1} C_4}{-{C_4}^{2} \left( C_5+\tau \right) ^{2}+16 e^{C_1} C_2}}}}}}
\end{eqnarray}

Note that, in principle, we must have  $C_2>0$ and $C_3>0$ since $\rho>0$.
Besides, if $C_4<0$  then $\tau<\sqrt{16 e^{C_1} C_2/C_4^{2}}-C_5$ or $C_4>0$ then $\tau>\sqrt{16 e^{C_1} C_2/C_4^{2}}-C_5$ in order to have real geodesic.

\begin{figure}[!htp]
	\centering	
	\includegraphics[width=6.5cm]{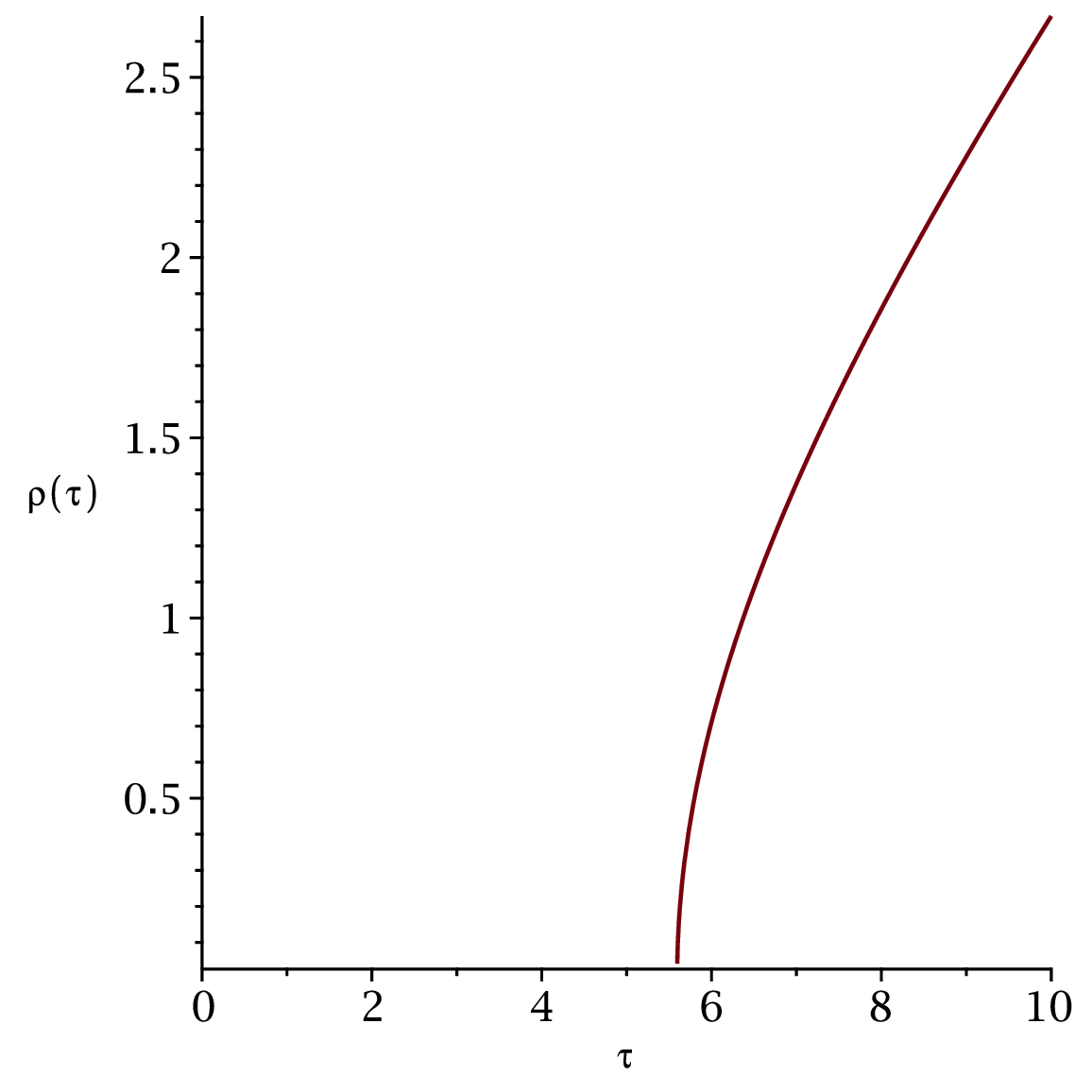}
	\includegraphics[width=6.5cm]{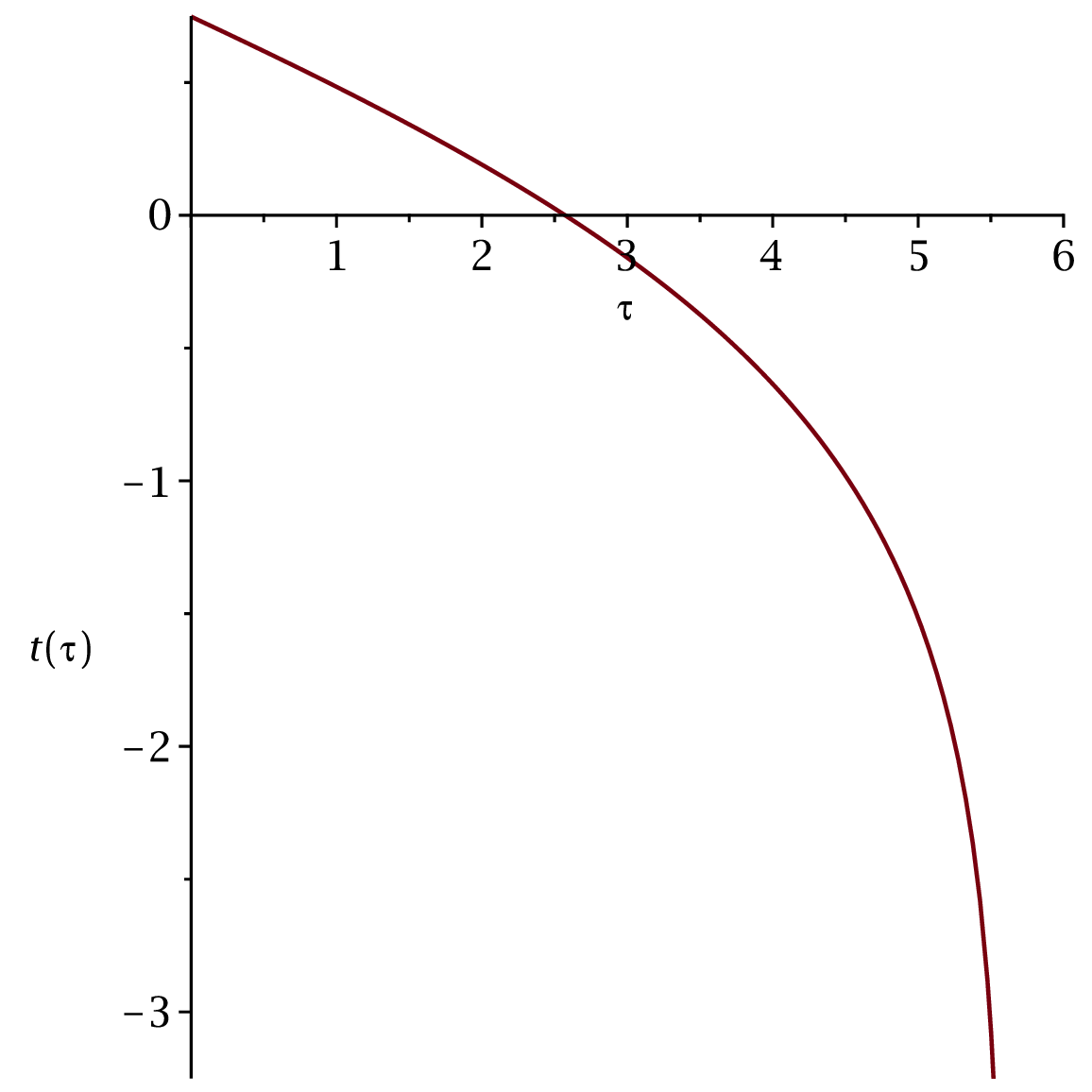}
	\includegraphics[width=6.5cm]{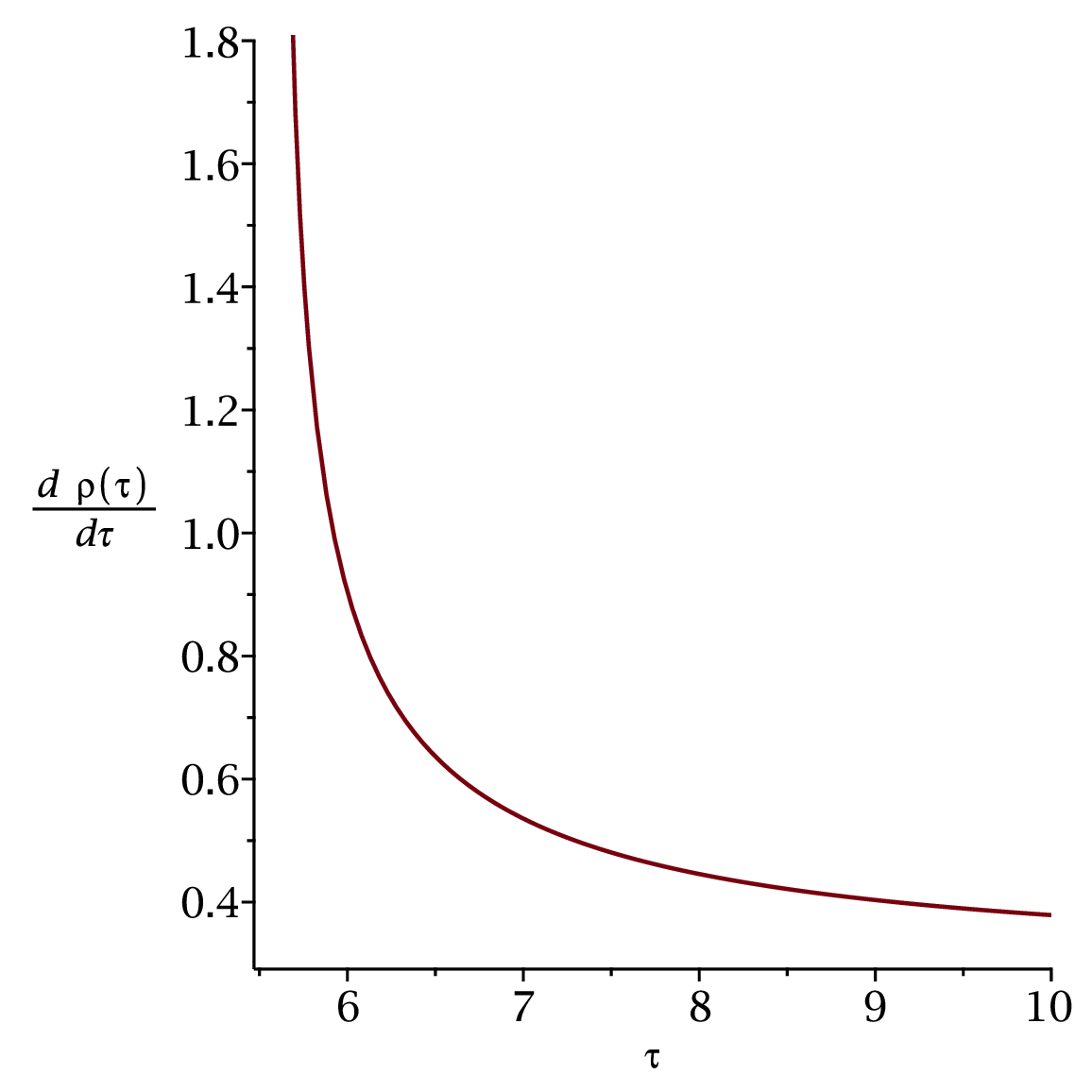}
	\includegraphics[width=6.5cm]{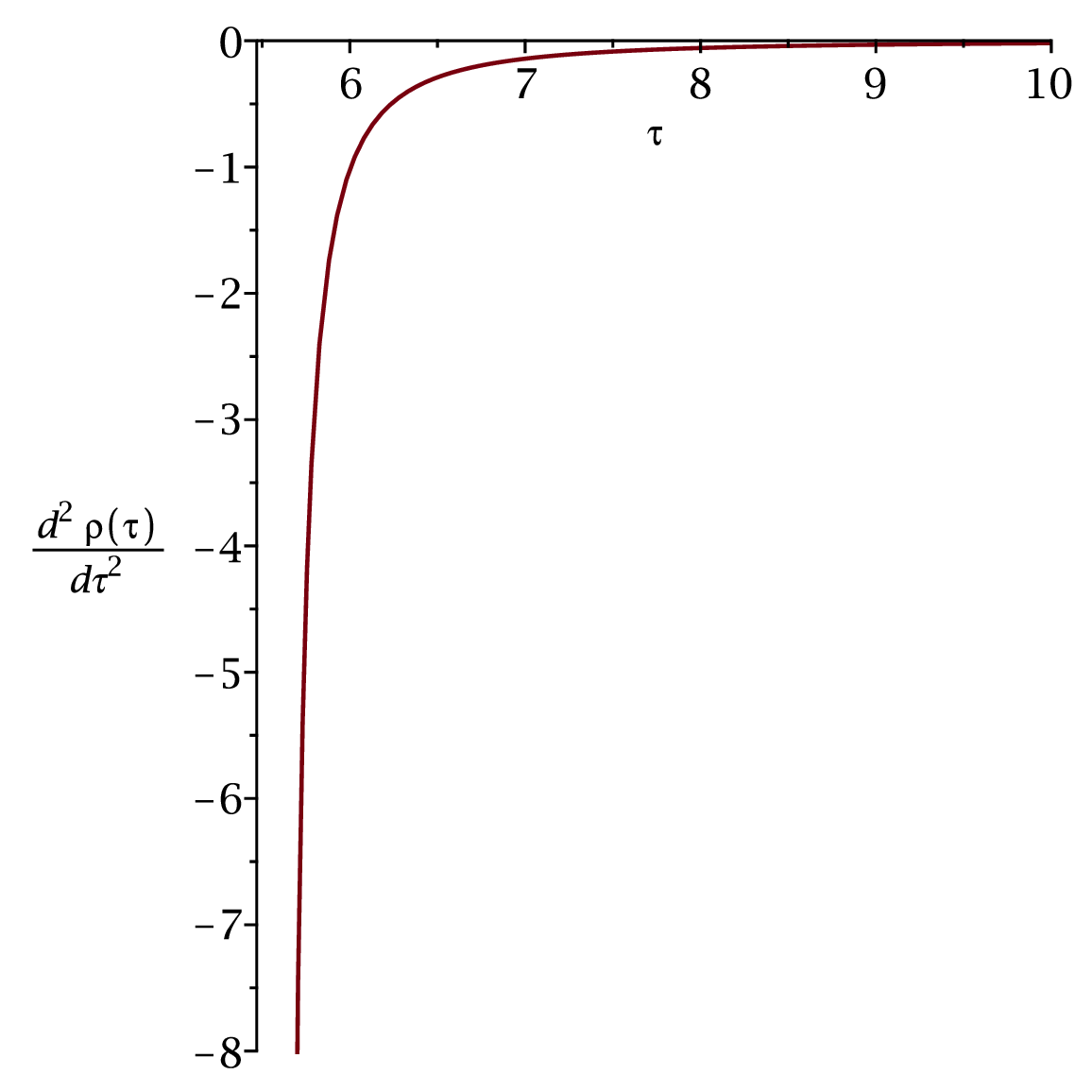}
	\caption{These figures show the time evolution of the $\rho$-geodesics of the Solution 5(a) and its velocity and acceleration of $\rho(\tau)$ on the plane ${z_0}=0$.
		We assume in these figures that $C_1=1$, $C_2=1$, $C_3=1$, $C_4=1$, $C_5=1$ and $C_6=1$.}
	\label{fig5a}
\end{figure}

See Figures \ref{fig5a}.

\section{Conclusions}
 
A procedure to find static axially symmetric solutions to the Einstein field equations is presented using particular conditions for the geodesics. 
We obtained two general solutions and five particular solutions,
which depend on the existence conditions for circular and $z$ direction motion.
Our aim consists making a thoroughrowly analysis of all the possible geodesics solutions stemming from this spacetime. In particular, the $z$ geodesic of the Solution 1 
presents the same physical characteristics
of relativistic jets highly energetic phenomena as described in \cite{Herrera2007}.
We have a positive acceleration for the test particle moving along $z$ direction 
at the beginning.
After sometime, we note that the test particle begins to decelerate. Finally, the
test particle stops the acceleration and it continues to travel at constant velocity.
In the paper, Herrera \& Santos, they have interpreted this initial acceleration due
to an existence of a repulsive force. However, in this work we can see that it is 
due only to the geometry of the spacetime.

\end{document}